\documentclass[12pt]{iopart}
\usepackage{graphicx}
\usepackage{xcolor}
\usepackage{hyperref}

\begin{document}

\title{Low energy neutrino-nucleus scattering in experiment and astrophysics}
\author{N Jachowicz, N Van Dessel and A Nikolakopoulos}
\address{Ghent University, Department of Physics and Astronomy, Proeftuinstraat 86, B-9000 Gent, Belgium}
\vspace{10pt}

\begin{abstract}
We review  the relevance of neutrino-nucleus interactions at energy transfers below 100 MeV for accelerator-based experiments, experiments at lower energies and for astrophysical neutrinos.  The impact  of low-energy scattering processes in the energy reconstruction analysis of oscillation experiments is investigated. We  discuss  the modeling of coherent scattering processes and compare its strength to that of inelastic interactions.  The presented results are obtained within a continuum random phase approximation approach.
\end{abstract}

%
%
%
%
%

\section{Introduction : low-energy neutrino-nucleus scattering}
Whereas accelerator-based neutrino-oscillation experiments \cite{Alvarez-Ruso:2017oui,Katori:2016yel} favor neutrino beams with energies peaking at a few  hundreds of MeVs \cite{miniboone,t2k,microboone} to several GeVs \cite{dune,argoneut,minerva,nova,dune,icarus}, the broad spectra with which neutrinos are generated in these experiments means that the signal in the detector is invariably the convolution of processes induced by neutrinos with energies ranging form very low to very high values. 
This necessitates the modeling of cross sections over a broad kinematic range with various energy and momentum transfers, including the low values we are considering here.
Moreover, nuclear responses essentially depend on energy and momentum transfer and not on incoming energy such that, regardless of the neutrino energy, the signal will contain  components stemming from processes with  low energy and momentum transfers.  In fact, forward lepton scattering events   favor reactions with low energy transfers to the nuclear system regardless of the incoming neutrino energy.  This means that  even experiments at relatively high energies have the potential to provide interesting information about e.g. supernova neutrinos and their interactions. Given the capabilities of Liquid Argon Time Projection Chamber (LArTPC) detectors to study more exclusive processes, the investigation specific kinematic ranges will gain in importance in future accelerator-based experiments.
 This obviously goes hand in hand with the need for an accurate theoretical description of these low-energy neutrino scattering processes.

 For the neutrino oscillation program in experiments such as e.g.~T2K \cite{MB:2010, T2K:2013}, MiniBooNE  \cite{miniboone}, MicroBooNE \cite{microboone} and DUNE \cite{dune} the neutrino nucleus cross section is the essential ingredient that connects the experimentally accessible variables to the neutrino energy that cannot be directly measured.  The cross section and its modeling hence provide the key to the reconstruction of the incoming  neutrino's energy.
The analyses in oscillation experiments relies on Monte Carlo event generators that are for the sake of computational efficiency often  restricted to using rather simple models, or simplified implementations of interaction models. The most commonly used model in oscillation analyses in the several hundreds of MeV region is the relativistic Fermi gas (RFG), in which the nucleus is described by its bulk properties. In Section \ref{reconstruction} we will discuss the relevance of adequate nuclear modeling for energy reconstruction in processes at low energies.

Over the past years much interest in the experimental community has gone to the differences of electron and muon neutrino-induced cross sections. This difference is important for the determination of the CP-violating phase and for the appearance of electron neutrinos in muon neutrino beams. 
Trivial differences between electron and muon neutrino interactions in charged-current scattering are contained in the lepton vertex, these differences are well understood.
For sufficiently large energy transfers and scattering angles the differences in kinematics due to the mass of the lepton become negligible and muon and electron cross sections are essentially the same.
For smaller scattering angles and  energy transfers however the mass of the outgoing lepton affects the kinematics to a greater extent, leading to muon neutrino interactions inducing a larger momentum transfer to the nuclear system than their electron neutrino counterparts. In this kinematic region the nuclear response is depleted by Pauli blocking and very sensitive to nuclear structure details, such that the large differences in momentum transfer can lead to significant differences in the responses for muon and electron neutrino induced interactions. The differences in the response can be so large that the $\nu_{\mu}$ induced cross section becomes larger than the $\nu_{e}$ one even though the latter is preferred by the lepton vertex. We will discuss these low-energy effects potentially affecting oscillation analyses and relevant for e.g. the MiniBooNE low-energy electron-like excess \cite{Aguilar-Arevalo:2018gpe} in Section \ref{nuemu}.

Apart from their interest in the axial structure of the nucleus and weak nuclear responses, experiments at low energies often focus on the weak response relevant for nuclei in stellar environments.  Although a lot of relevant  information can be obtained in e.g.~experimental beta decay studies, charge exchange (p,n) and (n,p) processes or (parity violating) electron scattering, neutrino scattering investigations are recognized as essential for a full understanding of the weak nuclear response that drives neutrino-nucleus scattering. 
Low-energy processes are important in   supernova dynamics and nucleosynthesis as well as for the terrestrial detection of supernova neutrinos \cite{Langanke:2019ggn}. 
Neutrino detection offers a window on the processes going on in the center of the star, whereas the more obvious optical observations are more limited to investigations of the stellar atmosphere.  With weak interactions as driving force in the stellar collapse, the importance of  neutrino interactions in these processes can hardly be overestimated.
As part of the collapse processes, large numbers of electron neutrinos are produced. These  leave the star unhindered until the core density of the star becomes sufficiently high for the neutrinos to become trapped \cite{Balasi:2015dba}.
After the core bounce, a burst of prompt neutrinos leaves the star core.  The remnant protoneutron star cools  as it emits the remaining  neutrinos  through pair production of neutrinos in all three generations.  Neutrino-wise, the net result of the supernova is the production and emission of ~10$^{57}$ neutrinos with individual energies up to a couple of tens of MeVs, collectively carrying away the energy that is released gravitationally in the supernova's collapse along with information about the processes going on in the core of the event.

Besides the details of the supernova process itself, the observation of neutrinos created therein is also of interest. This was first achieved at Kamiokande and IMB for the supernova SN1987A. Currently, several more detectors are operational, under development or being proposed \cite{dune,juno,hyperk}. Several of these involve the interaction of neutrinos with target atomic nuclei. This leads to the evident necessity of a thorough understanding of the scattering process between the (anti)neutrino and the nucleus and its energy dependence. One possible channel is through elastic interactions with nuclei, as was indeed recently achieved by the COHERENT collaboration for a CsI detector \cite{Akimov:2017ade}.

Apart from their contribution to the signal in accelerator-based experiments, cross sections induced by low-energy neutrinos have been measured for carbon and iron by dedicated experiments as LSND and KARMEN~\cite{Auerbach:2001hz,Maschuw:1998qh}. Several experiments have been proposed for the near future to measure supernova neutrinos at higher precision, including DUNE~\cite{dune} as well as e.g.~JUNO~\cite{juno} and Hyper-Kamiokande~\cite{hyperk}. These require significant research and development efforts in order to properly calibrate the detectors that are to be used. One necessity is to have access to neutrinos of similar kinematics to those produced in supernovas. The Spallation Neutrino Source at Oakridge (SNS)~\cite{Efremenko:2008an} is an example of such a facility. As a byproduct to its primary purpose of producing neutrons, the SNS also produces neutrinos of several flavors. These are born out of pions decaying at rest (DAR), yielding mono-energetic muon neutrinos at $E_{\nu_\mu} =$29.8 MeV as well as electron neutrinos and muon antineutrinos, the well-known Michel spectra with energies up to 52.8 MeV. One of the goals of the CAPTAIN~\cite{captain} program is to have the CAPTAIN Liquid Argon Time Projection Chamber (LarTPC) detector run at the SNS~\cite{Berns:2013usa}. This will yield crucial information on the neutrino-nucleus cross sections, but also on how to characterize these events in the analyses. A proper analysis of all the produced particles is a necessity to perform accurate calorimetry in SN neutrino experiments.

Whereas low-energy processes have a lot in common with scattering in the genuine quasi-elastic regime, some important differences have to be taken into consideration in their modeling.
Quasi-elastic scattering in the peak region is readily described within the impulse approximation and Fermi-gas based models usually do a fair job, especially for a description of inclusive processes.  For processes at low energy however, the response is more sensitive to nuclear structure details as e.g. binding energy, level schemes, and long-range correlations.
Moreover, a proper description of the wave function of the final-state nucleons and the related Pauli  blocking is absolutely essential.
These issues will be dealt with in the following paragraphs.
In literature, studies include calculations performed 
in RPA frameworks such as Refs.~\cite{McLaughlin:2004va,Engel:2002hg,Volpe:2001gy,Auerbach:1997ay,SUZUKI2003446,Bandyopadhyay:2016gkv,Cheoun:2011zza}, in a shell model~\cite{Hayes:1999ew,Kostensalo:2018kgh} (as well as hybrid models~\cite{Kolbe:2000np,Kolbe:1999vc}). Other models include the QRPA~\cite{Paar:2007fi,Samana:2008pt} and local Fermi Gas-based RPA approaches~\cite{Nieves:2004wx,SajjadAthar:2005ke,Singh:1998md}. Previous applications of the CRPA framework are 
 found in Refs.~\cite{Kolbe:1999au,Jachowicz:2002hz}. A comparison between shell model, RPA and CRPA results was performed in~\cite{Volpe:2000zn}. 
Model-independent calculations of cross sections were done in Refs.~\cite{Pourkaviani:1990et,Fukugita:1988hg}

\section{Cross sections}
In the low-energy limit, the two vertices between which the gauge boson propagates  are not resolved, 
and the weak processes can be described as point interactions mediated by a contact force. Within this effective theory, the weak interaction Hamiltonian $\widehat{H}_W$ then obtains the current-current structure~:
\begin{equation}
\widehat{H}_W= \frac{G_F}{\sqrt 2} 
\int d\vec{x}\:\;\;  \hat{\jmath}_{\mu,leptonic}(\vec{x})\;\;  \widehat{J}^{\mu,hadronic}(\vec{x})  , \label{ham}
\end{equation}
with 
 $G_F$ the weak coupling constant, which  for charged current interactions the coupling constant has to be multiplied by the factor $\cos\theta_C$ in order to take  Cabibbo mixing into account.

For sufficiently low momentum transfers, only lowest-order contributions to the hadronic current have to be retained \cite{walecka,fetterwalecka1971}.  In a standard non-relativistic expansion, the different contributions to the nucleon current read as~: 
\begin{eqnarray} 
\vec{J}^{\alpha}_V \left( \vec{x} \right) 
&=&\vec{J}^{\alpha}_{convection} \left( \vec{x} \right) +
   \vec{J}^{\alpha}_{magnetization} \left( \vec{x} \right) \nonumber \\
 \mbox{with}  & &  \vec{J}^{\alpha}_{c}\left(\vec{x}\right)=
   \frac{1}{2Mi} \sum_{i=1}^{A} G_E^{i,\alpha} \left[ \delta \left(\vec{x}
   -\vec{x}_i\right) \stackrel{\rightarrow}{\nabla}_i-
   \stackrel{\leftarrow}{\nabla}_i \delta \left(\vec{x}
   -\vec{x}_i\right)\right],  \nonumber \\
   & &\vec{J}_{m}^{\alpha} \left(\vec{x}\right)= \frac{1}{2M}\: 
    \sum_{i=1}^{A} \,G_M^{i,\alpha}
   \stackrel{\rightarrow}{\nabla}\:
    \times
   \: \vec{\sigma}_i \: \delta \left(\vec{x} -\vec{x}_i\right),
   \label{nonrel1}\\ 
   \vec{J}_A^{\alpha}\left(\vec{x}\right)&=& \sum_{i=1}^A G_A^{i,\alpha}\: 
   \vec{\sigma}_i\:
   \delta \left(\vec{x} -\vec{x}_i\right),  \label{nonrel2} 
\end{eqnarray}
\vspace*{-.5cm}
\begin{eqnarray}
J_V^{0,\alpha}\left(\vec{x}\right)=\rho_V^{\alpha}\left(\vec{x}\right)&=&
 \sum_{i=1}^{A} G_E^{i,\alpha}\: \delta \left(\vec{x} -\vec{x}_i\right), 
 \label{nonrel3} \\
J_A^{0,\alpha}\left(\vec{x}\right)=\rho_A^{\alpha}\left(\vec{x}\right)&=& 
\frac{1}{2Mi}\; \sum_{i=1}^{A}\, G_A^{i,\alpha} \; \vec{\sigma}_i \cdot
\left[ \delta \left(\vec{x}
 -\vec{x}_i\right) \stackrel{\rightarrow}{\nabla}_i- 
   \stackrel{\leftarrow}{\nabla}_i \delta \left(\vec{x}
   -\vec{x}_i\right)\right].\nonumber\\ \label{nonrel4} \\
J_P^{0,\alpha}\left(\vec{x}\right)=\rho_P^{\alpha}\left(\vec{x}\right)&=& 
\frac{m_{\mu}}{2M }\; \sum_{i=1}^{A}\, G_P^{i,\alpha} \;\vec{\nabla}\cdot \vec{\sigma}_i \;\delta \left(\vec{x}-\vec{x}_i\right), \label{nonrel5} 
\end{eqnarray}
where  the summations extend over all nucleons in the nucleus, the index $\alpha$ identifying the isospin character of the contribution.  The form factors in these expressions have to be attributed different values for charged and neutral current interactions.  The last expression (\ref{nonrel5}) corresponds to the pseudoscalar contribution.  At threshold, this coupling  can be shown to be  proportional to the mass of the outgoing lepton, and   hence results in very small contributions to the cross section \cite{walecka}.  

The inclusive double differential cross-section  for scattering of a neutrino with energy $E_i$ becomes :
\begin{eqnarray}\label{eindelijk}
\frac{\mathrm{d}^3\sigma}{\mathrm{d}E_f \mathrm{d}^2\Omega_f} &=
\frac{G_F^2 E_f k_f}{\pi}  ,\nonumber \\
&\times \left[v_{CC}W_{CC} + v_{CL} W_{CL} + v_{LL} W_{LL} + v_T W_T + hv_{T'} 
W_{T'}\right],
\end{eqnarray}
where $E_f$ and $k_f$ are the the outgoing lepton's energy and momentum. $h$ is $+$ or $-$ for neutrinos and antineutrinos, encoding the influence of the leptons helicity on the cross section.

\begin{eqnarray}
v_{CC} &= 1 + \zeta \cos{\theta}  ,\nonumber \\
v_{CL} &= -\left( \frac{\omega}{q}(1 + \zeta \cos{\theta}) + \frac{m_f^2}{E_f q} \right) ,\nonumber \\
v_{LL} &= 1 + \zeta \cos{\theta} - \frac{2E_i E_f}{q^2}\zeta^2\sin^2{\theta}  ,\nonumber \\
v_{T} &= 1 - \zeta \cos{\theta} + \frac{E_i E_f}{q^2}\zeta^2\sin^2{\theta} ,\nonumber \\
v_{T'} &= \frac{E_i + E_f}{q}(1 - \zeta \cos{\theta}) - \frac{m_f^2}{E_f q}  ,
\end{eqnarray}
with $\zeta = \frac{k_f}{E_f}$ and $\theta$ the lepton's scattering angle, $m_f$ the final lepton's mass,
\begin{eqnarray}
W_{CC} &=  \sum_{J \geq 0} \sum_{l,j,j_h} \left| \langle \Phi_f || \widehat{\mathcal{M}}_{J}(q) || \Phi_0 \rangle \right|^2,\nonumber \\
W_{CL} &=  -2 \sum_{J \geq 0} \sum_{l,j,j_h} \mathrm{Re} \left[ \langle \Phi_f 
| \widehat{\mathcal{J}}^{el}_{J}(q) || \Phi_0 \rangle \left(\langle \Phi_f || 
\widehat{\mathcal{J}}^{mag}_{J}(q) || \Phi_0 \rangle \right)^* \right],
\nonumber \\
W_{LL} &=  \sum_{J \geq 0} \sum_{l,j,j_h} \left| \langle \Phi_f || \widehat{\mathcal{L}}_{J}(q) || \Phi_0 \rangle \right|^2,\nonumber \\
W_{T} &=  \sum_{J \geq 1} \sum_{l,j,j_h} \left( \left| \langle \Phi_f || \widehat{\mathcal{J}}^{el}_{J}(q) || \Phi_0 \rangle \right|^2 + \left| \langle \Phi_f
 || \widehat{\mathcal{J}}^{mag}_{J}(q) || \Phi_0 \rangle \right|^2 \right),\nonumber \\
W_{T'} &=  2 \sum_{J \geq 1} \sum_{l,j,j_h} \mathrm{Re} \left[ \langle \Phi_f|| \widehat{\mathcal{J}}^{el}_{J}(q) || \Phi_0 \rangle \left(\langle \Phi_f || 
  \widehat{\mathcal{J}}^{mag}_{J}(q) || \Phi_0 \rangle \right)^* \right].
\label{vgl9}
\end{eqnarray}
In these expressions, $l$ and $j$ denote the orbital and total momentum quantum numbers of the outgoing particles, whereas $j_h$ denotes the quantum number of the hole state. The energy transfer is denoted by $\omega$, momentum transfer is $q$.
The multipole operators are defined as :
\begin{eqnarray}
\widehat{\cal M}_{JM} \left( q \right) &= &\int d\vec{x} \, \left[
\jmath_J\left( qr \right) Y_J^M \left(\Omega_x\right) \right]\:
\widehat{J}_o\left(\vec{x}\right),\nonumber \\
\widehat{\cal L}_{JM}\left(q\right) &=& \frac{i}{q}
\int d\vec{x} \left[ \vec{\nabla} \left( \jmath_J\left(q r\right)
Y_J^M\left(\Omega_x \right)\right) \right] \cdot \widehat{\vec{J}}
\left(\vec{x}\right), \nonumber \\
\widehat{\cal J}^{el}_{JM} \left( q \right) &=& \frac{1}{q}
\int d\vec{x} \left[ \vec{\nabla} \times \left( \jmath_J\left(q r\right)
\vec{\cal Y}_{J,J}^{M} \left(\Omega_x
\right)\right)\right] \cdot \widehat{\vec{J}}\left(\vec{x} \right), \nonumber \\
\widehat{\cal J }^{mag}_{JM} \left( q \right) &=& \int d\vec{x}
\left[ \jmath_J\left(q r \right) \vec{\cal Y}^M_{J,J}\left(\Omega_x\right)
\right] \cdot \widehat{\vec{J}} \left(\vec{x}\right).\label{opera}
\end{eqnarray}
Here, $\widehat{\cal M}_{JM}$ and $\widehat{\cal L}_{JM}$ denote the Coulomb and longitudinal operator respectively, whereas  $\widehat{\cal J}^{el}_{JM}$ and
 $\widehat{\cal J }^{mag}_{JM}$ are the transverse electric and magnetic operators.  For each multipole transition $J^{\pi}$ only one part -vector or axial vector- of an operator is contributing.  From the expression (\ref{vgl9}) it is clear that $J=0$ transitions are suppressed due to the lack of a transverse contribution in these channels.  Still, neutrinos are able to excite $0^-$ states in nuclei, whereas electrons cannot.  The second and third part of the expression show that for inclusive processes, there is interference between the Coulomb and the longitudinal (CL) terms and between both transverse contributions, but not between transverse and CL terms.   From the angular dependence of the kinematic factors, it is clear that for backward lepton scattering mainly transverse terms contribute, while for $\theta=0$ CL-contributions dominate.

For charged current neutrino scattering reactions, the outgoing particle is a charged lepton.  In this case, the wave function of the outgoing lepton should be described by the scattering solutions for the outgoing particle in the Coulomb potential generated by the final nucleus.  Theoretically, this could be achieved by expanding the outgoing lepton wave into distorted partial waves calculated in this Coulomb potential. In practice, one makes use of effective schemes.  At low outgoing lepton energies, of the order of magnitude applicable to e.g.~beta decay, one can make use of the Fermi function~\cite{Engel:1997fy} to account for Coulomb distortion :

\begin{equation}
\zeta(Z',E_f)^2 = 2(1+\gamma_0)(2k_f R)^{-2(1-\gamma_0)} \frac{|\Gamma(\gamma_0+i\eta)|^2}{(\Gamma(2\gamma_0+1))^2},
\end{equation}
where $R\approx 1.2 A^{1/3} \mathrm{fm}$ is the nuclear radius, $\gamma_0=\sqrt{1-(\alpha Z')^2}$, $E_f$ is the outgoing lepton's energy, $k_f$ the outgoing momentum and $\eta=\pm \frac{\alpha Z' E_f}{k_f}$. with $+$ and $-$ for neutrinos and antineutrinos respectively. Similarly, the final nuclear charge $Z'$ is equal to $Z+1$ or $Z-1$ for $\nu$/$\bar{\nu}$, respectively.  Finally, $\alpha$ is the fine structure constant, equal to $\frac{e^2}{4\pi\epsilon_0} \approx \frac{1}{137.036}$.  This approximation assumes that the outgoing leptons only contribute sizable through the s-wave component to the reaction strength, and is therefore not applicable at higher outgoing lepton energies~\cite{Engel:1997fy}. Therefore, in these regimes, we  consider a different scheme, the modified effective momentum approximation, detailed in Ref.~\cite{Engel:1997fy}, where the energy and momentum of the final lepton is shifted to an effective value by the Coulomb energy in the center of the nucleus that is assumed to have a uniform density :
\begin{equation}
E_{eff} = E_f - V_c(0) = E_f \pm \frac{3}{2}\frac{Z' \alpha  }{R},
\end{equation}
which induces a factor in the cross section that accounts for the change in phase space:
\begin{equation}
\zeta(Z',E_f)^2 = \frac{E_{eff} k_{eff} }{E_f k_f},
\end{equation}
as well as a shift in the momentum transfer $q \rightarrow q_{eff}$ in the calculation of the transition amplitudes. In practice, one can interpolate between these two schemes. This consists simply of taking for each value of $\omega$ the value that is closest to unity. The effect of the Coulomb interaction is to increase or decrease both the differential and the integrated cross section for neutrinos and antineutrinos, respectively. 

Another topic of importance to low-energy calculations is the  '$g_A$ problem'. For free nucleons, the static value of $G_A(Q^2=0) \equiv g_A \approx 1.28$. It is the axial part of the current that at low energies is responsible for generating the Gamow-Teller (GT) transition strengths. It was noticed by Chou et. al~\cite{Chou:1993zz} that in the context of $\beta$-decay calculations, a shell-model framework overpredicts this strength. Consequently, shell model calculations often make use of an effective 'quenched' axial coupling constant in order to reproduce experimental GT strengths. This quenched value is not unambiguous, as is noted in Refs.~\cite{Suhonen:2018ykq,Engel:2016xgb}. A possible solution to this problem was recently suggested in \cite{Pastore:2017uwc}, where it was posited that the need for such a quenching may appear as a consequence of an incomplete treatment of the nuclear wave functions through either the absence of correlations or an insufficiently large model space. Indeed, their quantum Monte Carlo calculations show satisfactory agreement with experiment for light nuclei with $A = 6 - 10$ when using the free axial coupling, showing the importance of an accurate treatment of correlations in the nucleus and providing insight into a possible solution to the long-standing $g_A$ problem.  In this work we will use the free value of $g_A$.

\section{Long-range correlations, collective excitations and RPA}
The nuclear force is strong and gives
rise to a variety of  phenomena as pairing of the nucleons, collective excitations, nuclear
vibration and rotation, ...  These mainly show up at low energies,  where the full richness of the nuclear dynamics becomes manifest.
Ab-initio methods \cite{Lovato:2015qka} are beyond doubt the optimum choice for the description of inclusive interactions off light 
nuclei, but computational limitations necessitate the use
 of approximate schemes for heavier nuclei and higher energies.
 \begin{figure}
   \begin{center}
\includegraphics[width=.6 \textwidth,scale=0.5]{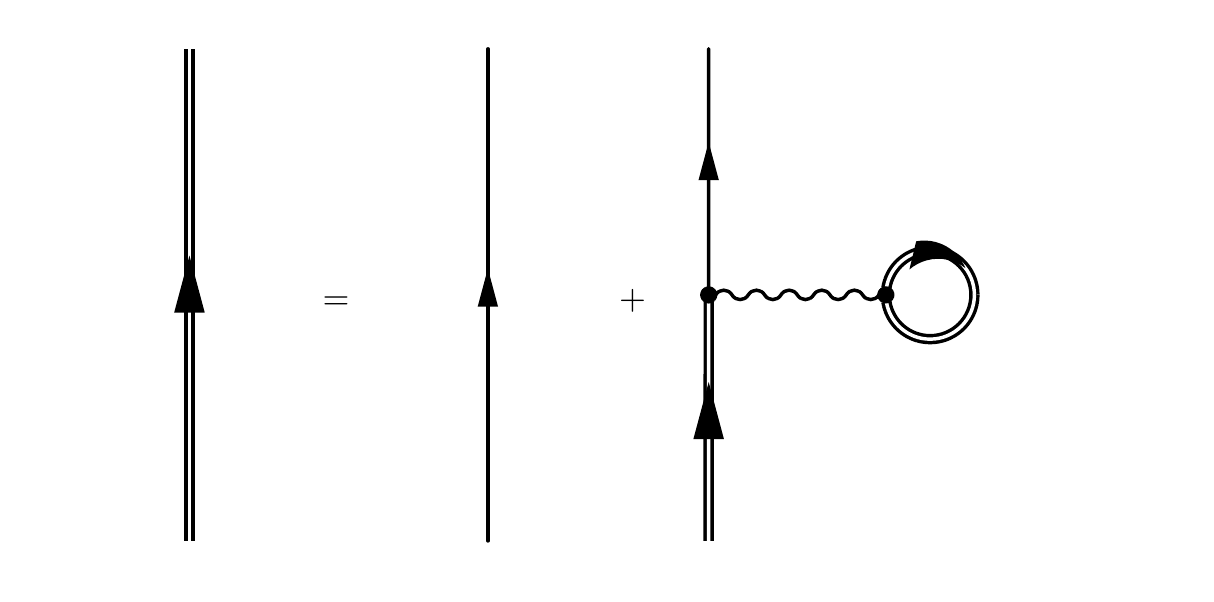}
\caption{The mean-field approach as expressed in a Green's function formalism. The single-particle propagator is dressed by iteration of all first-order interactions.}\label{mf}
   \end{center}
\end{figure}
A mean field description of the nucleus, as depicted diagrammatically in Fig.~\ref{mf}, already captures a lot of the dynamics that ties the nucleons together in the nucleus, and provides in general an adequate description of genuine quasi-elastic cross sections in the region of the QE peak, where the higher order contributions of the residual interaction will only induce small corrections.  At smaller energies however, collective effects and long-range correlations play a crucial role and strongly influence the nuclear response to electroweak probes.
\begin{figure}
   \begin{center}
\includegraphics[width=.7\textwidth,scale=0.5]{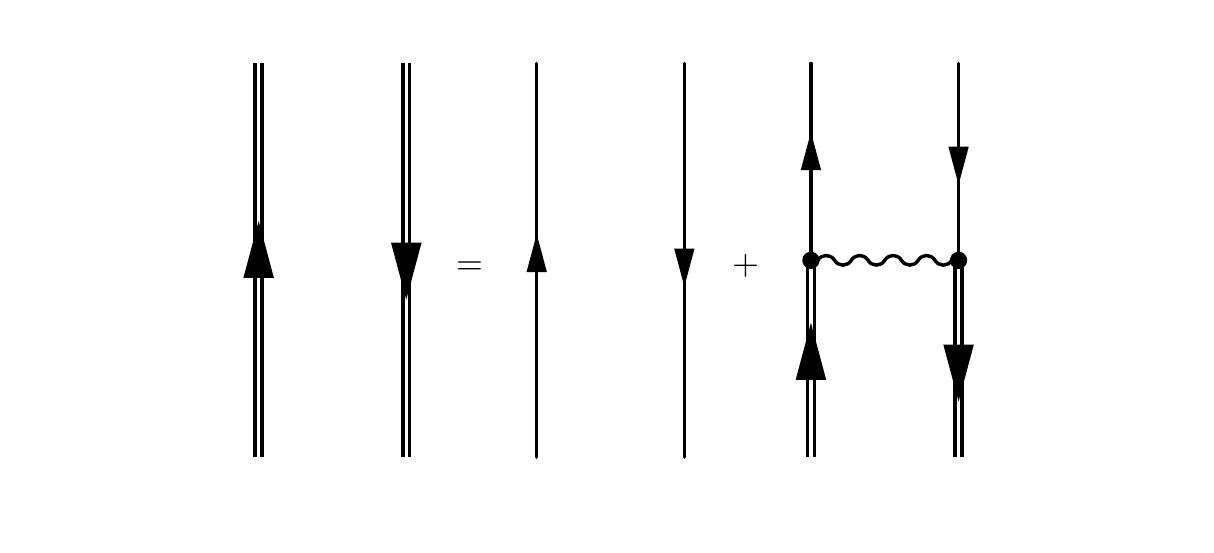}
\caption{The Random Phase Approximation as expressed in a Green's function formalism.}\label{rpa}
   \end{center}
\end{figure}

Generally, a   shell-model approach to the nuclear many-body problem relies on
the separation of the nuclear Hamiltonian in a well-known mean-field contribution,
diagonal in the considered single-particle basis and a residual term, accounting for
the correlations between            nucleons.
The residual interaction will bring along a mixing of the mean field
model states and realistic eigenstates of the nuclear system will be linear combinations
of these basis wave functions.
The single-particle wave-functions are used to construct a basis of many-body
wave-functions for the nucleus. These then serve to set up the Hamiltonian matrix
for the nuclear many-body system. This matrix is diagonalized in order to obtain
eigenvalues and the corresponding eigenstates of the nucleus.
A major problem using this approach is the dimension of the matrices to be
diagonalized, rapidly growing with increasing model sizes.

Confronted with this drawback, a number of approximations have been designed,
focusing on various aspects of the problem. Next to the
Hartree-Fock approximation, considering only mean-field properties of
the problem, more elaborate techniques as e.g.~random phase approximation (RPA) approaches, shown diagrammatically in Fig.~\ref{rpa},  were developed.
Contrary to mean-field descriptions where a nucleon experiences the
presence of the others only through the mean field generated by their
mutual interactions, the RPA  allows
correlations to be present even in the ground state of the nuclear
system and additionally allows the particles to interact by means of
the residual two-body force.  The random phase approximation goes
 one step beyond the zeroth-order mean-field approach and  describes a
nuclear state as the coherent superposition of particle-hole
contributions.
\begin{equation}
\left| \Psi_{RPA} \right\rangle = \sum_c \, \left\{\, X_{(\Psi,C)} \;\left| ph^
{-1}\right\rangle\: -\: Y_{(\Psi,C)} \;\left| hp^{-1}\right\rangle \, \right\} 
.\label{xyrpa}
\end{equation}
The summation index $C$ stands for all quantum numbers defining a
 reaction channel unambiguously~:
\begin{equation}
C=\{n_h,l_h,j_h,m_h,\varepsilon_h;l_p,j_p,m_p,\tau_z\},\label{C}
\end{equation}
where the  indices $p$ and $h$ indicate whether the considered quantum numbers 
relate to the particle or the hole state,  $\varepsilon_h$ denotes the binding-energy of the hole state and $\tau_z$ defines the isospin character of the particle-hole pair.
General excited states are obtained as linear combinations of these
particle-hole configurations. As the RPA approach describes nuclear
excitations as the coherent superposition of individual particle-hole
states out of a correlated ground state, this approach allows to account for some of the collectivity present in
the nucleus, especially the long-range correlations important in a low-energy regime.

 In our  approach,  the unperturbed
wave-functions are generated with a
Hartree-Fock (HF) calculation using the SkE2 Skyrme ~\cite{Waroquier:1986mj} force parametrization to build the single-nucleon potential.  The potential $\hat{V}$ of the nuclear force is modeled a as sum of two-- and three--nucleon interactions:
\begin{equation}\label{eq:potential}
\hat{V} = \sum_{i<j} \hat{V}^{(2)}_{ij} + \sum_{i<j<k} \hat{V}^{(3)}_{ijk},
\end{equation}
where these interactions are 
\begin{eqnarray}
\hat{V}^{(2)}_{ij} = t_0(1+x_0 P_\sigma) \delta^{(3)}(\vec{r}_i-\vec{r}_j) \nonumber \\
 + \frac{1}{2}t_1 \left( \delta^{(3)}(\vec{r}_i-\vec{r}_j) k_{ij}^2 + k_{ij}^{'2} \delta^{(3)}(\vec{r}_i-\vec{r}_j) \right) \nonumber \\
 + t_2 \overleftarrow{k}'_{ij}\cdot \delta^{(3)}(\vec{r}_i-\vec{r}_j) \overrightarrow{k}_{ij} \nonumber \\
 + iW_0\left( \vec{\sigma}_i + \vec{\sigma}_j \right) \cdot \left( \overleftarrow{k}'_{ij}\times \delta^{(3)}(\vec{r}_i-\vec{r}_j) \overrightarrow{k}_{ij} \right) \nonumber \\
 + \frac{1}{6}t_3(1-x_3)(1+ P_\sigma) \rho\left(\frac{\vec{r}_i+\vec{r}_j}{2}\right)\delta^{(3)}(\vec{r}_i-\vec{r}_j) \nonumber \\
 + V^{Coul}_{ij} \label{intint}
\end{eqnarray}
and 
\begin{eqnarray}
\hat{V}^{(3)}_{ijk} = x_3 t_3 \delta^{(3)}(\vec{r}_i-\vec{r}_j) \delta^{(3)}(\vec{r}_i-\vec{r}_k) \nonumber \\
+ \frac{1}{6} t_4 \left[ (k_{ij}^{'2} + k_{jk}^{'2} + k_{ki}^{'2}) \delta^{(3)}(\vec{r}_i-\vec{r}_j) \delta^{(3)}(\vec{r}_i-\vec{r}_k) \right. \nonumber \\
+ \left. \delta^{(3)}(\vec{r}_i-\vec{r}_j) \delta^{(3)}(\vec{r}_i-\vec{r}_k) (k_{ij}^{2} + k_{jk}^{2} + k_{ki}^{2}) \right]. 
\end{eqnarray}
Here $P_\sigma$ is the spin--exchange operator$\frac{1}{2}\left( 1 + \vec{\sigma}_i \cdot \vec{\sigma}_j \right)$, along with the momentum operators $\overrightarrow{k}_{ij} = \frac{-i}{2} (\overrightarrow{\nabla}_i-\overrightarrow{\nabla}_j)$ and $\overleftarrow{k}'_{ij} = \frac{i}{2} (\overleftarrow{\nabla}_i-\overleftarrow{\nabla}_j)$, acting on the right and the left respectively. The same Skyrme parameterization is used as residual nucleon-nucleon interactions in the RPA calculation.  This approach makes
self-consistent HF-RPA calculations possible \cite{Jachowicz:1998fn,Jachowicz:2002rr}.

Solving the equations for the RPA  polarization propagator $\Pi^{RPA}$
\begin{eqnarray}
  \Pi^{RPA}(x_1,x_2;E)=& \Pi^{0}(x_1,x_2;E)\nonumber\\
  &+\frac{1}{\hbar}\int dxdx' \Pi^{0}(x_1,x;E)V(x,x')\Pi^{RPA}(x',x_2;E),
\end{eqnarray}
in coordinate space, allows one to treat the energy 
continuum in an exact way, hence the name continuum RPA (CRPA).  In the above $\Pi^{0}$ is the mean-field propagator, and $V$ the antisymmetrized interaction.  This equation is the basis for the calculation of   the CRPA transition matrix elements needed to evaluate \ref{vgl9}.  The reduced CRPA transition densities are determined by a set of coupled integral equations :
\begin{eqnarray}
\left\langle\Psi_0||X_{\eta J}||\Psi_C(J;E)\right\rangle_r\;\;&\;\;=\;\;&\;\;-\: \left\langle h||X_{\eta J}||p\right\rangle_r \nonumber \\&&\hspace{-4.3cm} \;\;+\;\;\;\; \sum_{\mu,\nu} \,\int \!dr_1\!\! \int \!dr_2\;\: U_{\mu\nu}^{J}(r_1,r_2)\;\, {\cal R}\, \left(R_{\eta\mu; J}^{(0)} (r,r_1;E)\right) \;\;\left\langle \Psi_0||X_{\nu J}||\Psi_C(J;E)\right\rangle_{r_2},\nonumber \\
\label{transdens}
\end{eqnarray}
where $\Psi_0$ and $\Psi _C$ denote the ground and excited-state CRPA wave functions,   $r$ indicates the radial dependence of the quantity.  The sum runs over all included reaction channels.  The $X_{\eta JM}$ correspond to the different terms in the interaction \ref{intint}, $U$ contains the radial dependence of the interaction.
 The unperturbed radial response functions are defined as
\begin{equation}
  \int dr \int dr'\:\: R_{\eta\mu ;JM}^{(0)}(r,r';E)=\frac{1}{\hbar}\int dx\int dx'\; \;\,X_{\eta JM}(x)\;\; \Pi^{(0)}(x,x';\omega)\;\;X^{\dagger}_{\eta' JM}(x')\;.\label{radres}\end{equation}

The importance of an accurate treatment of this collectivity through long-range correlations lies in the need to model giant resonance excitations of nuclei, which contribute non-trivially to the reaction strength at low neutrino energies. This stands in contrast with higher (few 100s of MeV) energies, where the quasielastic peak, corresponding to an electroweak probe interacting with a quasifree single nucleon, can be accurately described in a mean-field approach. It should be noted however, that even at higher energies, collectivity is important for events of low energy and momentum transfer, $\omega$ and $q$. Giant resonances (GRs) are formally described as coherent particle-hole excitations induced by the nuclear current operator. Different parts of this operator are responsible for different types of GRs, characterized by different selection rules. An example of such a resonance is the isovector giant dipole resonance (IVGDR), which corresponds with $\Delta L = 1$, $\Delta S = 0$ and $\Delta T = 1$. The macroscopic interpretation of this excitation is the oscillation of the protons in the nucleus against the neutrons. Other types of GR exist, as covered in e.g.~\cite{harakeh}. While the CRPA formalism is capable of predicting the position and strength of collective excitations, it does not predict the width of the resonance accurately. This is because within the CRPA, the configuration space is limited to $1p1h$ excitations with fixed hole energies. The finite width results from coupling to higher--order configurations.  To remedy this, in our approach, the width of the states is taken into account in an effective way by folding the responses with a Lorentzian with an effective width of 3 MeV.

An important benchmark for any model describing neutrino-nucleus interactions is a comparison with the much more abundant electron-scattering data.
For the CRPA approach this has been done successfully in \cite{Pandey:2014tza}.

\section{Cross sections at low energies}
\begin{figure}
   \begin{center}
  \includegraphics[height=.3\textheight,scale=0.4]{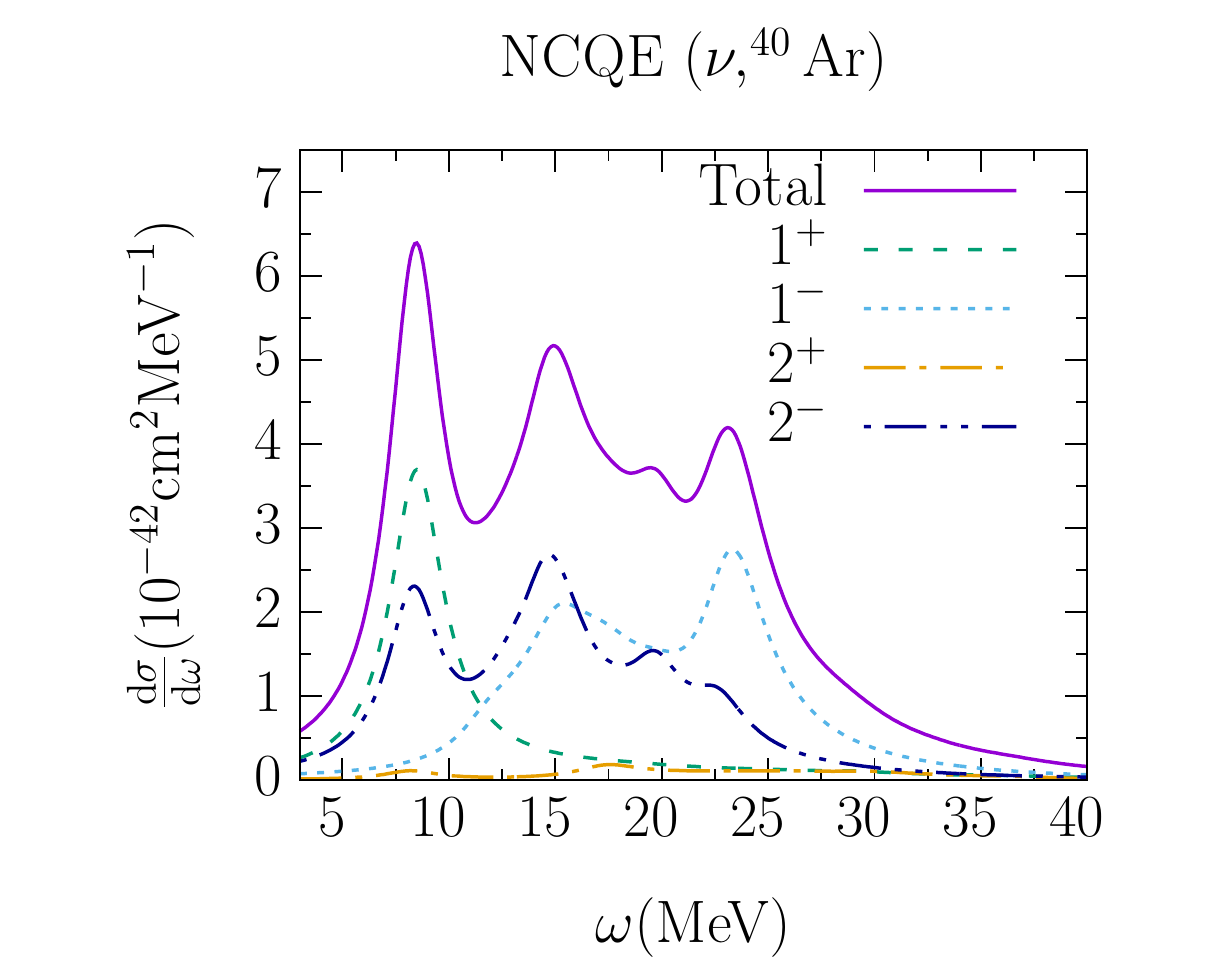}
\caption{Differential cross section for the neutral current reaction $^{40}Ar+\
\nu_{ 50\:MeV}\rightarrow \:^{40}\!Ar^*+\nu'$ and its dominant multipole contributions.}
\label{een}
   \end{center}
\end{figure}
Figure \ref{een}  shows the differential
cross section for neutral scattering of a 50 MeV neutrino off $^{40}$Ar as a function of the excitation energy
$\omega$ of the nucleus, and its most important multipole contributions.  The plot shows the typical features of low-energy neutrino-nucleus interactions.
The differential neutrino scattering cross-sections are  typically of the order of 10$^{-4 2}$ cm$^2$ per MeV.  The influence of single-nucleon levels is apparent in the lower part of the spectrum. At slightly higher energy transfers, the strength stemming from  giant resonance contributions shows up.
The figure shows that even at these very low energies, forbidden transitions provide 
a considerable contribution to the total interaction strength.

\begin{figure}
   \begin{center}
\includegraphics[width=.7\textwidth,scale=0.6]{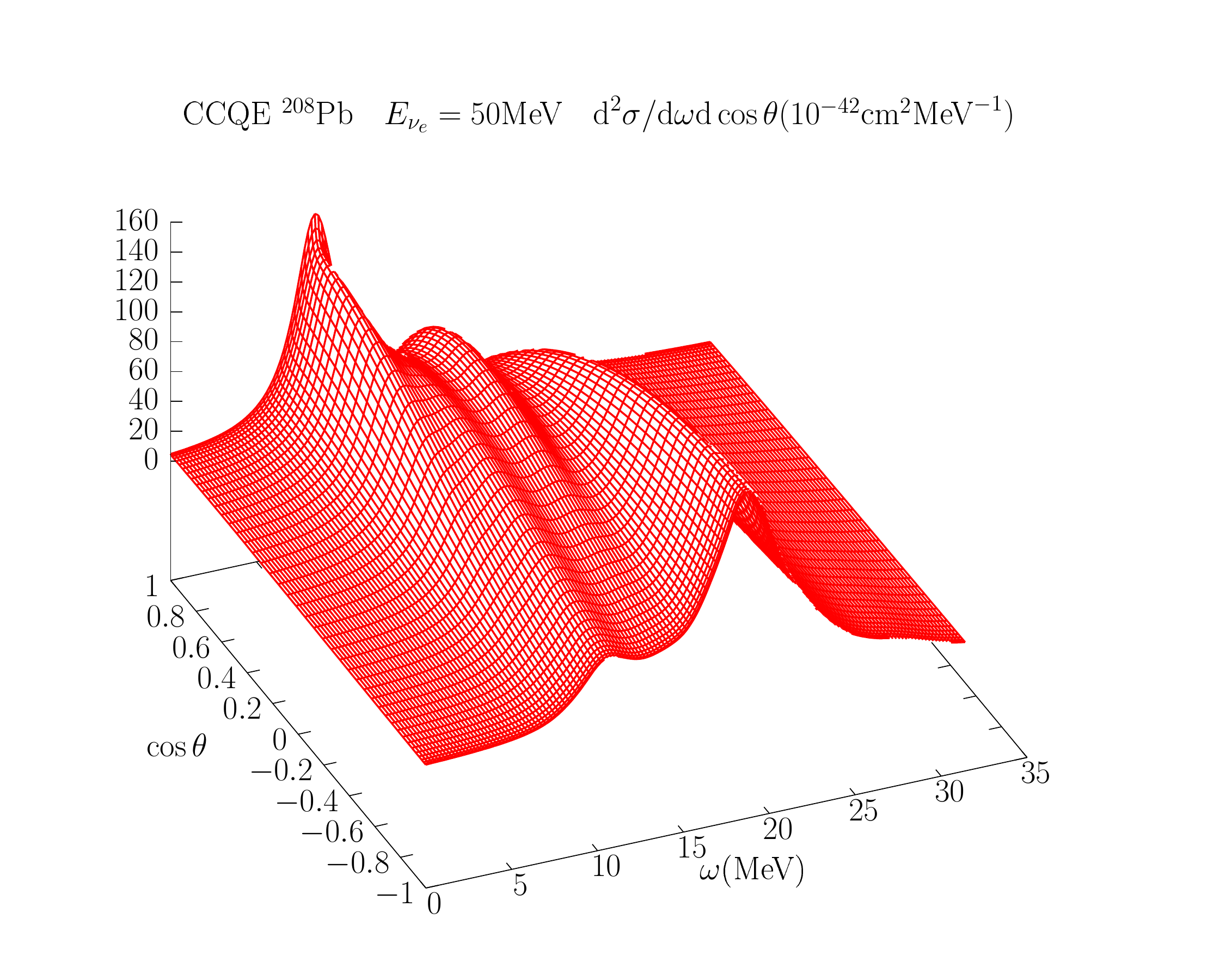}
\caption{Double differential cross-section for charged current scattering of 50\
  MeV  neutrinos  off $^{208}$Pb, $\theta$ represents the scattering angle of the lepton.}
\label{Pbdd}
   \end{center}
\end{figure}

Figure \ref{Pbdd} shows double differential cross sections for charged-current scattering of a 50 MeV electron neutrino off $^{208}$Pb.  The importance of back-scattering channels for the final charged electrons is a most striking feature of this process, that  becomes even  more outspoken for lighter nuclei.

\begin{figure}
   \begin{center}
\includegraphics[height=.3\textheight,scale=0.4]{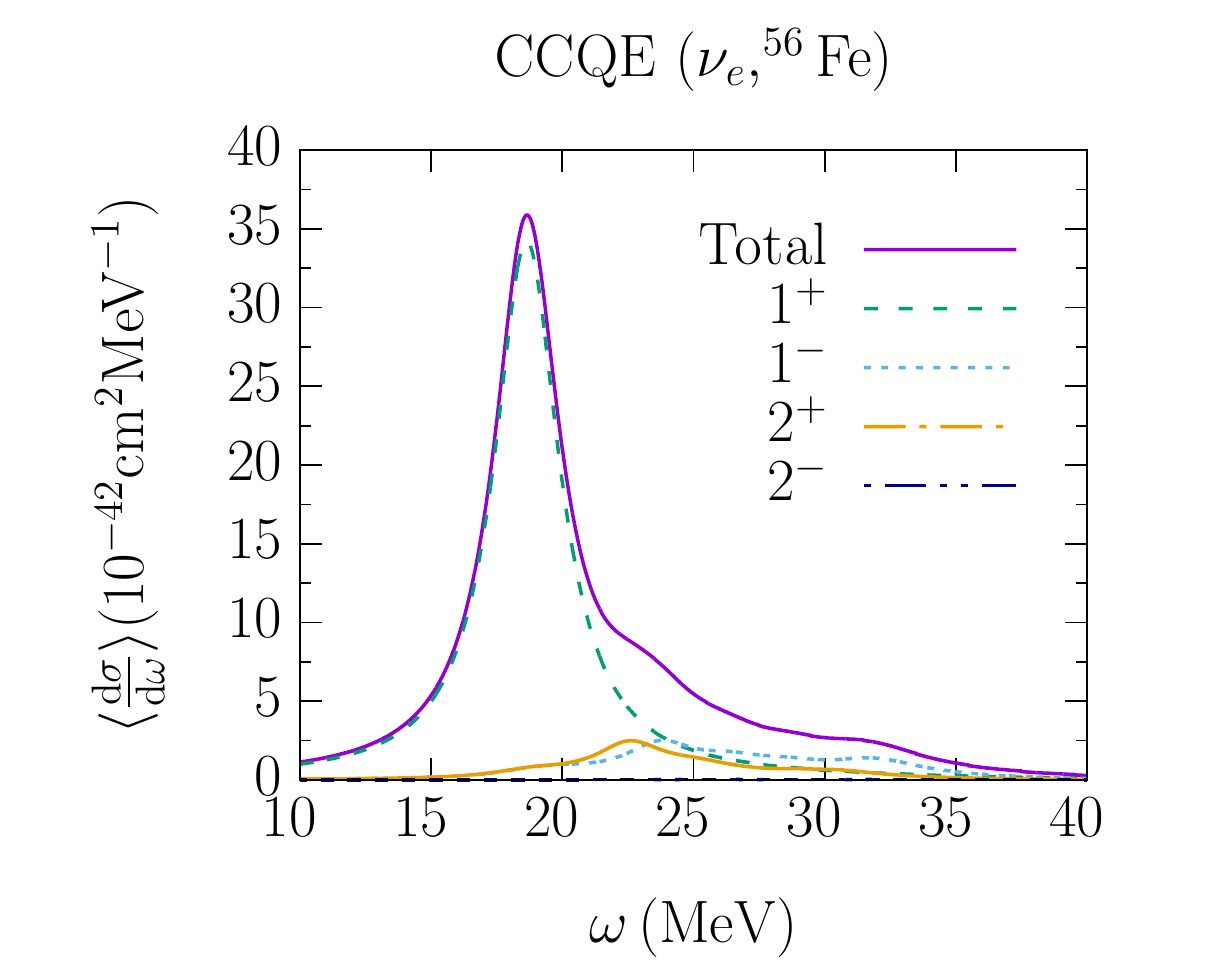}
\caption{Flux--folded differential cross section for pion decay-at-rest neutrinos scattering off $^{56}$Fe, $\omega$ represents the excitation energy. The main multipole contributions are shown separately.}
  \label{darfe} \end{center}
\end{figure}

\begin{figure}
   \begin{center}
\includegraphics[height=.3\textheight,scale=0.4]{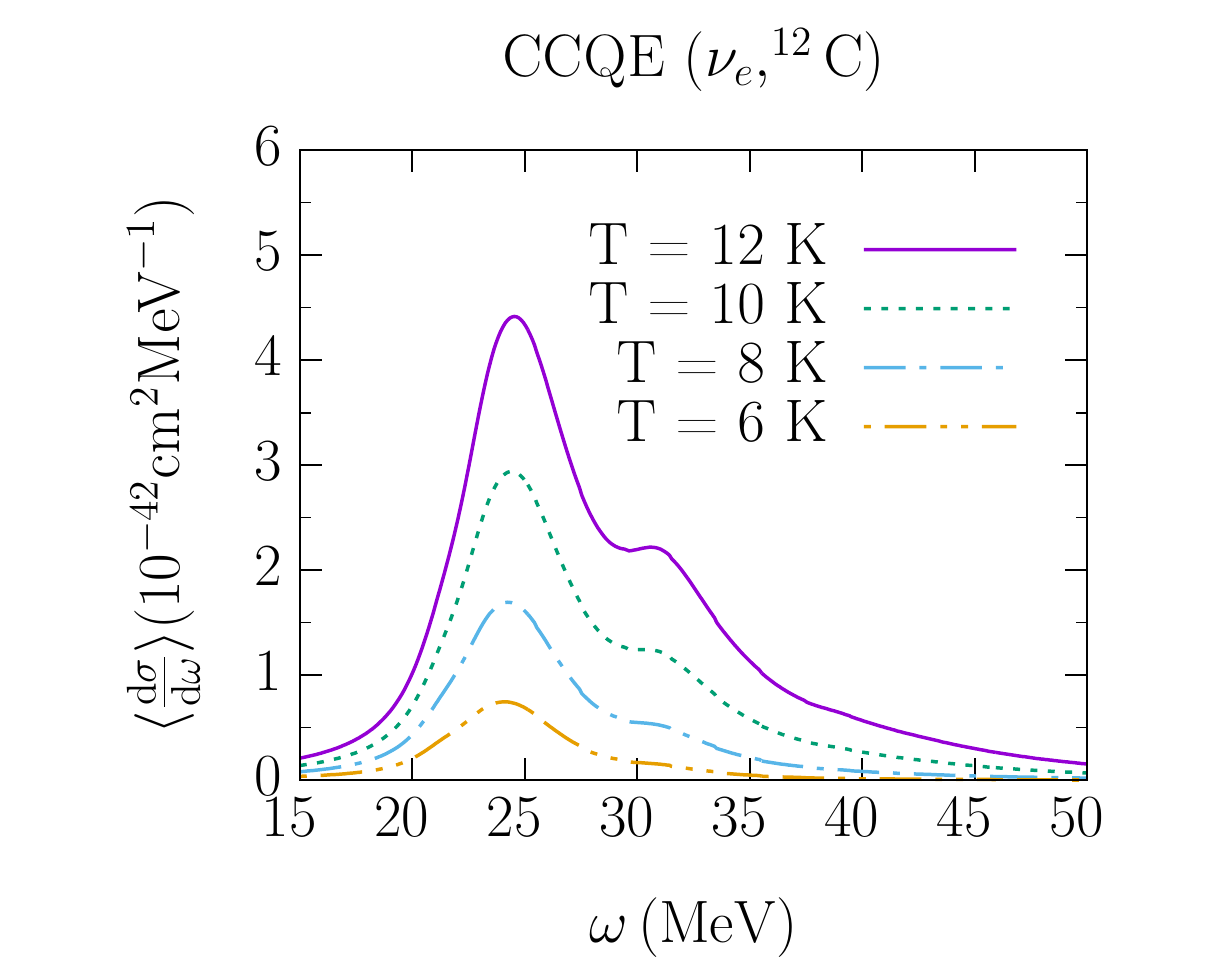}
\caption{Flux-folded differential cross section for neutrinos in a Fermi-Dirac spectrum at various temperatures scattering off $^{12}$C. $\omega$ represents\
 the excitation energy.}
  \label{fermic} \end{center}
\end{figure}

In Fig. \ref{darfe}, we show the single differential cross section for charged-current scattering of pion decay-at-rest (DAR) electron neutrinos off $^{56}$Fe.
 While the Gamow-Teller ($1^+$) strength of the $1^+$ transition dominates the total strength,  the  $1^-$
$2^+$ transitions are also important for a full modeling of the cross 
section, especially at slightly higher energies. We also show figure \ref{fermic}, with the single differential cross section for charged-current scattering of electron neutrinos distributed according to a Fermi-Dirac spectrum, at various temperatures, off $^{12}$C.  Most striking is the strong dependence of the cross section strength on the temperature of the supernova-neutrino energy-spectrum \cite{Jachowicz:2003iz}.

Whereas the features described above are obvious when dealing with cross sections induced by neutrinos with small energies,  they are also present in reactions with higher incoming energies \cite{Pandey:2016jju}.  Even for energies up to  $\sim$1 GeV and higher, for forward lepton scattering reactions, the energy transfer to the nucleus is small, and low energy aspects of the nuclear response will dominate the cross section.

\begin{figure}
   \begin{center}
\includegraphics[width=.7\textwidth,scale=0.6]{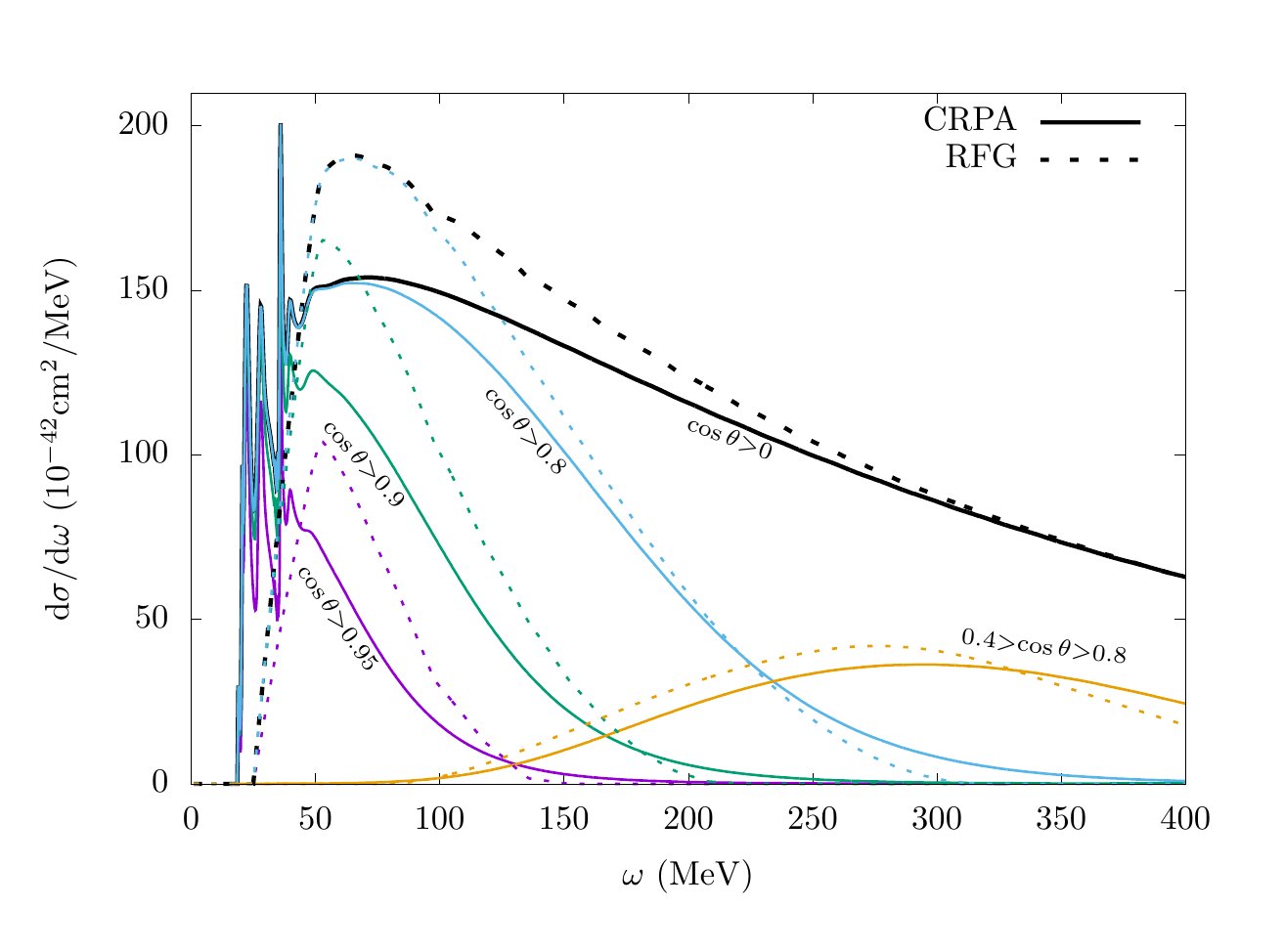}
\caption{Differential cross-section for charged-current scattering off carbon as a function of the energy transfer at an incoming energy of $E_\nu = 1~
{GeV}$ for $\cos\theta > 0$.
Contributions from different lepton scattering angles are shown separately}
\label{fig:RFG_CRPA_angles}
   \end{center}
\end{figure}

Figure~\ref{fig:RFG_CRPA_angles} shows the single differential cross section for an incoming energy of $1~\mathrm{GeV}$ where contributions from different angular regions are shown separately.  For leptons scattering angles with $\cos\theta>0.8$, the influence of nuclear structure details and giant resonances is obvious.
Comparing with the predictions of an relativistic Fermi gas  model, we see that for forward scattering angles the proper treatment of the nuclear dynamics and Pauli blocking is essential. In the RFG model the low $\omega$ region is strongly depleted by Pauli blocking, while the strength is shifted towards larger values.

Worth mentioning in this context is the recent measurement of kaon decay-at-rest (KDAR) neutrino cross sections by the MiniBooNe collaboration \cite{Aguilar-Arevalo:2018ylq}.  These have the double merit that they are monochromatic with an energy of 236 MeV and moreover result in cross sections in the transition region between the low-energy and the genuine quasielastic regime.

\section{Low-energy processes and reconstruction}\label{reconstruction}
In the analysis of oscillation experiments, the reconstruction of the impinging neutrino's energy is essential.  Neutrino-nucleus interactions are used to count the number of neutrinos of a certain flavor  in the near and far detector, but in order to extract oscillation information from these data, the energy of the neutrino is equally important.  The broad energy distribution of neutrino beams and experimental limitations in measuring vertex energy necessitate to resort to reconstruction procedures.
\begin{figure}
   \begin{center}
\includegraphics[width=0.5\textwidth]{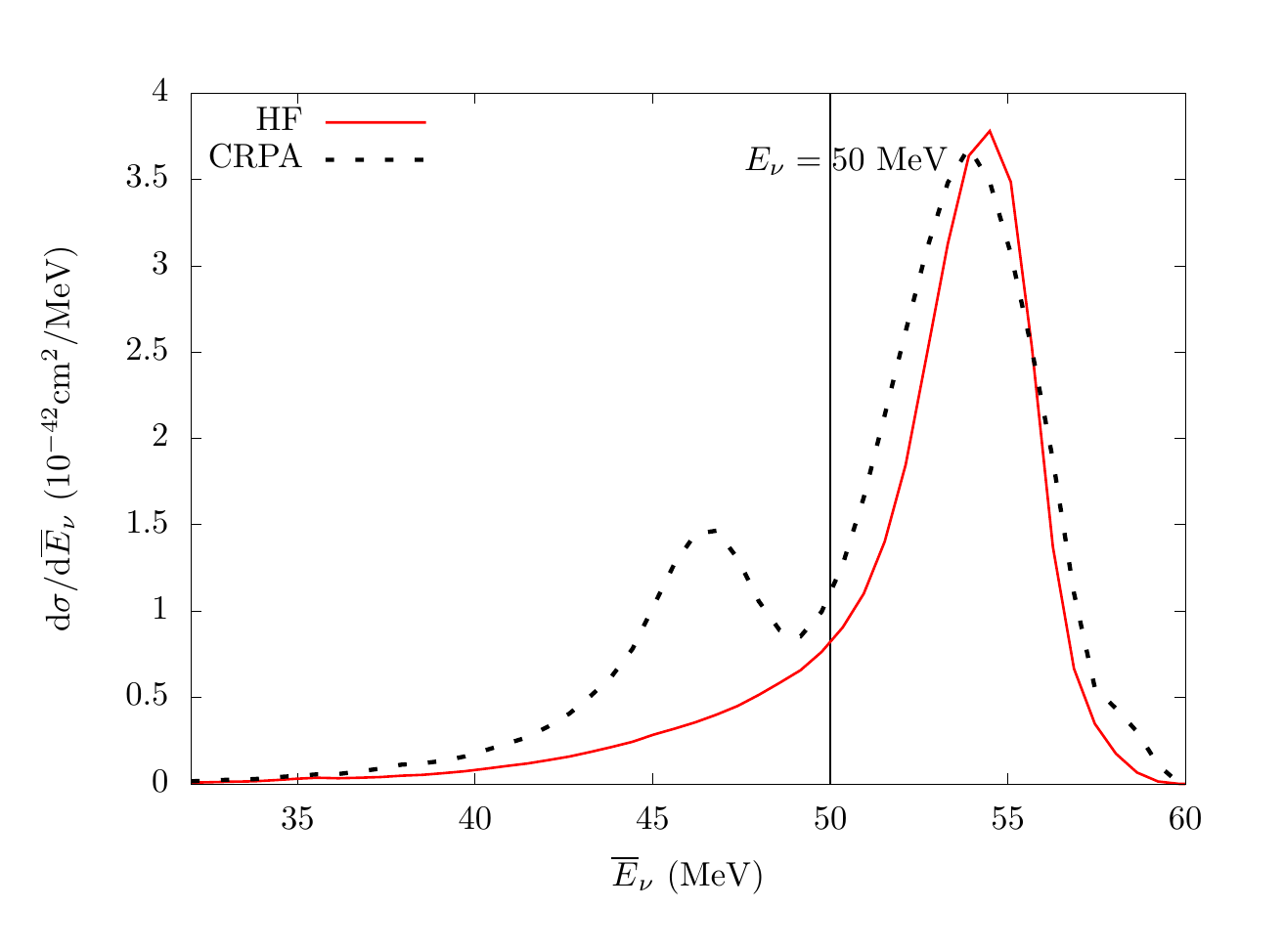}
\caption{HF and CRPA charged-current differential cross sections off $^{12}$ C in terms of reconstructed energies with $E_B = 25~\mathrm{MeV}$ for an incoming neutrino energy of $50~\mathrm{MeV}$.}
\label{fig:d_Erec}
   \end{center}
\end{figure}

The approach  used in the experimental analyses of MiniBooNE and T2K \cite{MB:2010, T2K:2013} is a reconstruction procedure based  on the kinematics of the charged final state lepton
\begin{equation}
\label{eq:erecbind}
\overline{E}_{\nu}= \frac{2M^{\prime}_nE_{l}-({M^{\prime}_n}^2 
+m_l^2 -M_p^2)}{2(M^{\prime}_n - E_l +P_l\cos\theta)},
\end{equation}
where $M_n^{\prime}=M_n-E_B$ is the adjusted neutron mass, with $M_n$($M_p$) the neutron(proton) rest mass, and $E_B$ a fixed binding energy.
The lepton rest mass and momentum are $m_l$ and $P_l$ respectively. This approach assumes scattering off a static nucleon, the reconstructed energy only corrected for a fixed binding.
The simplicity of this procedure, largely omitting the influence of the nucleon's motion and spreading in the nuclear medium, leads to a distribution of reconstructed energies around the true neutrino energy that has to be provided by a model for the neutrino-nucleus interaction.

The model dependency of these distributions for neutrino interactions with incoming energies of several MeVs has been studied within various approaches, often focusing on the inclusion of reaction mechanisms which contribute to the QE experimental signal if the reaction products are not detected or misidentified.
 These interactions include multi-nucleon emission, pion production and re-absorption, ~\cite{Martini2012,GiBUU:2p2h,Nieves:Unfolding,LeitnerandMosel,Lalakulich:QErec,VanCuyck:2016fab,VanCuyck:2017wfn,Martini:recooscill, Ericson:Reco}.
The effects of final-state interactions (FSI) on the reconstructed energy distributions for purely quasielastic interactions have been studied with a spectral function approach in Ref.~\cite{Ankowski:reco}, and with the CRPA in Ref.~\cite{Nikolakopoulos:2018sbo}.

At low energies, the reconstructed energy distribution is significantly distorted by nuclear structure details and giant resonance contributions.
In Fig.~\ref{fig:d_Erec}, we show the single differential cross section in terms of reconstructed energies, taking $E_B=25~\mathrm{MeV}$, for an incoming energy of 50 MeV.  The $E_B$ value corresponds to the weighted average binding energy of single-particle energies in carbon.
A striking feature is that giant resonances clearly contribute in the low $\overline{E}_{\nu}$ region, while the strength in the vicinity of the  peak  is slightly reduced.
Another important fact is that the distribution is strongly offset from the true neutrino energy. This is due to the shape of the cross section at low $\omega$, where the cross section peak is located before the naive peak position inferred from Eq.~(\ref{eq:erecbind}).

The difference in peak position between the true and reconstructed energy distributions could removed by treating $E_B$ as a free parameter. This however is not actually the goal of a reconstruction analysis. The trivial shift induced by varying $E_B$ is merely a redefinition of the $\overline{E}_\nu$. The most important aspect in this analysis, is the model dependence of the reconstructed distribution, as it is only through an interaction model that one can link the reconstructed to the real energy distribution.

\begin{figure}
   \begin{center}
\includegraphics[width=0.5\textwidth, scale=0.1]{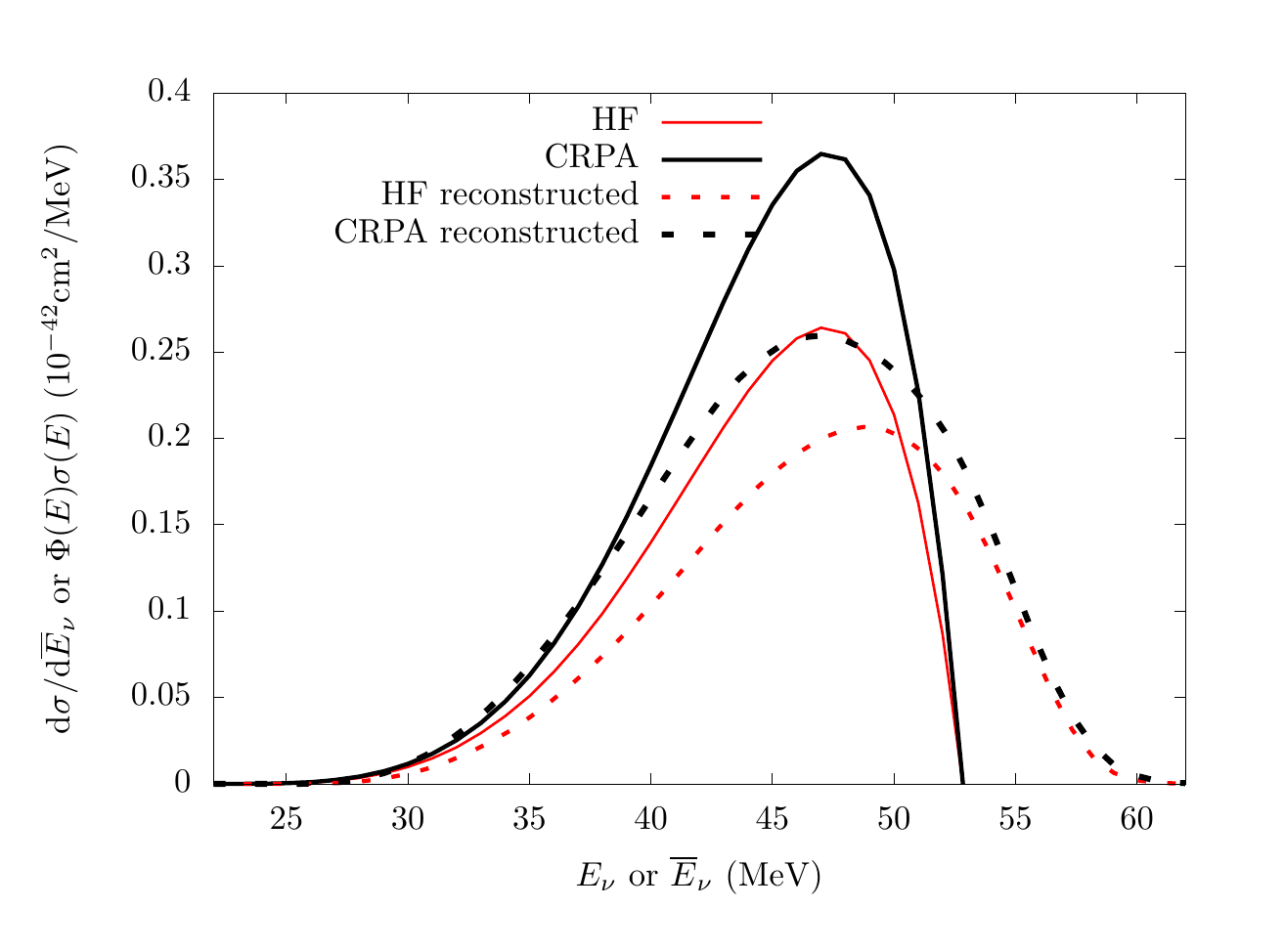}
\caption{The cross section for charged-current scattering off $^{12}$C in terms of reconstructed energies averaged over the $\pi$DAR flux, is compared to the total cross section in both the HF and CRPA models.}
   \label{fig:PidarRec}\end{center}
\end{figure}
The overall effect of this on an experimentally accessible distribution is shown in Fig.~\ref{fig:PidarRec}, where the cross section is averaged over the $\pi$DAR electron neutrino flux.
Even though experiments dedicated to the study of DAR neutrino interactions, such as CAPTAIN running at the Spallation Neutron Source (SNS)~\cite{captain}, often make use of reconstructing the decay products, the reconstruction from lepton kinematics still provides interesting insights as it is essentially a clean kinematic variable which does not depend on the details of the hadronic final state, and is 
not affected by  misidentification or non-detection of decay products.
We see that the GR contribute considerably to the total cross section. The reconstructed distributions are spread  over a large region, the CRPA peak is shifted to the left and the low $\overline{E}_\nu$ tail is enhanced compared to a HF approach. This is readily understood as the GR contribution which indeed 
contributes for small reconstructed energies as seen in Fig.~\ref{fig:PidarRec}.

\begin{figure}
   \begin{center}
\includegraphics[width=0.5\textwidth, scale=0.1]{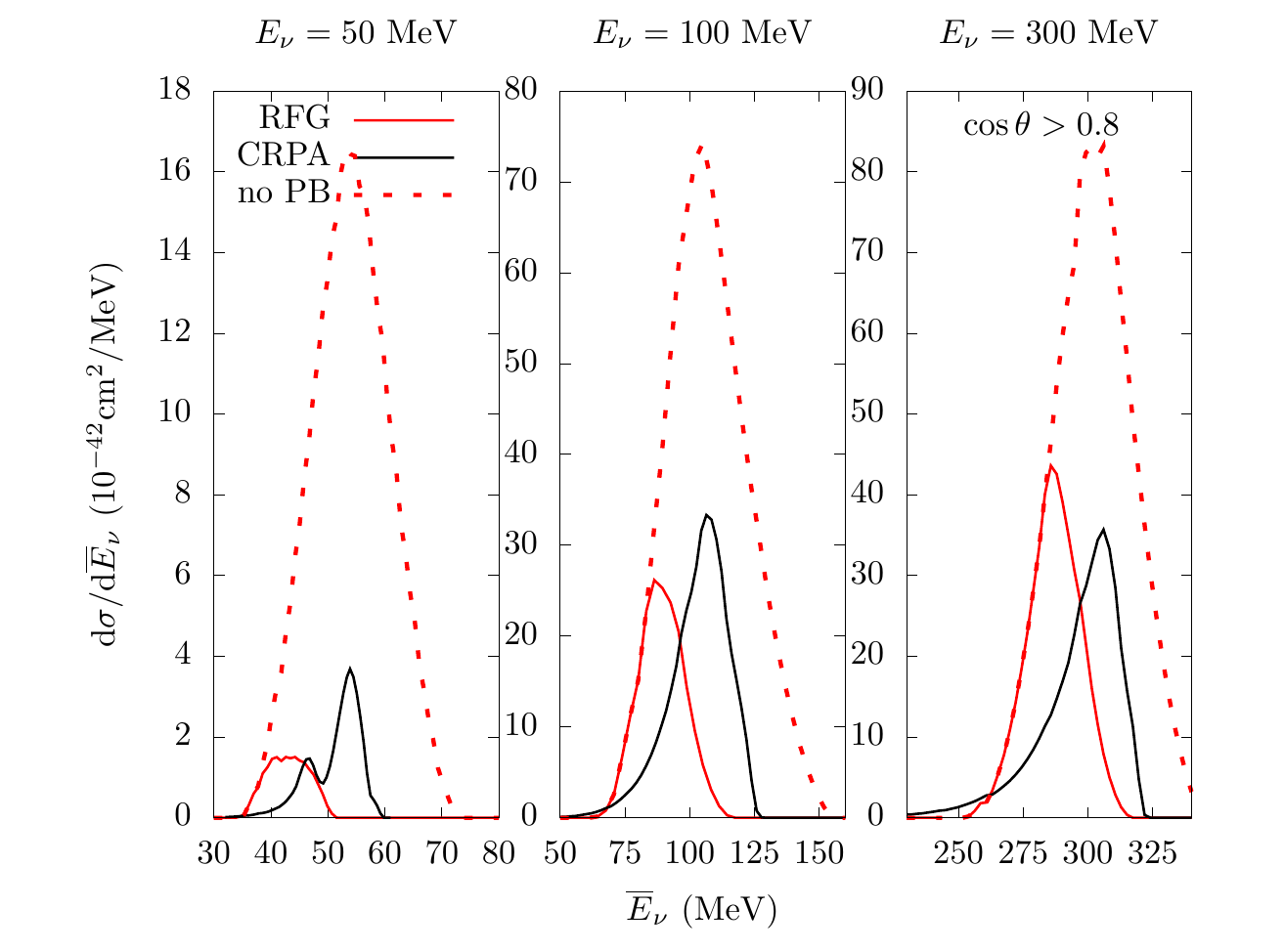}
\caption{The single differential cross section in terms of reconstructed energies in the RFG and CRPA approaches. We also show the RFG results where Pauli blocking is neglected and only the energy shift is taken into account. In the rightmost panel only contributions for forward scattering angles are shown, while in the other panels the whole phase space is considered.}
   \label{fig:PB_RFG}\end{center}
\end{figure}

These findings also relate to another important aspect of the description of  the low energy region in neutrino-nucleus cross sections.  In our approach final state nucleon wave functions are evaluated in the nuclear potential, naturally including Pauli blocking and elastic distortion of the final nucleon's wave function.  Pauli blocking is often treated naively in approaches that do not involve a mean field description for the nucleus. In a commonly used  approach, a model not including elastic FSI, will set the cross section to zero if the outgoing nucleon's momentum is smaller than the Fermi momentum. 
A such approach generally results in a rather good description of the magnitude of the interaction cross section for inclusive kinematics, however the shape of the cross section is strongly offset. 
  In a Pauli blocked Fermi gas the low $\omega$ region is completely depleted, with a far too large cross section at higher energy transfers. The effect on the reconstructed distribution is shown in Fig.~\ref{fig:PB_RFG}, where the RFG results with and without Pauli blocking are compared to the CRPA cross sections.
Clearly Pauli blocking is necessary to obtain a  reasonable overall strength for   the cross
 section. However, removing all contributions with the outgoing nucleon momentum  below the Fermi level leads to  cross section distributions which strongly differ from the 
CRPA results.
The same effect is present for larger incoming energies if we restrict 
the cross section to forward scattering angles as shown e.g. in  the rightmost panel of Fig.~\ref{fig:PB_RFG}.

\section{Electron versus muon neutrino induced processes}\label{nuemu}
As accelerator-based experiments rely on counting neutrinos in their near and far detectors and base the oscillation analysis on different numbers between the former and the latter, a thorough understanding of cross sections differences between muon and electron neutrinos is essential.  For neutral-current processes, lepton universality predicts equal interaction strengths, but for charged-current processes the mass of the final lepton influences the size of the cross sections.
At higher energies, the differences are small and result in slightly larger cross sections for electron-neutrino induced processes, with the lighter lepton in the final state.  For low energies however, the situation is  less clear. 
In this regime, the nuclear responses are strongly sensitive to the nuclear structure and  dynamics and this is reflected in a strong model dependence in the outcome of various calculations \cite{Ankowski:2017yvm,Martini:2016eec}.  The treatment of the final nucleon state and especially a proper description of Pauli blocking is important as shown in~\cite{Nikolakopoulos:2019qcr}.

\begin{figure}
   \begin{center}
\includegraphics[width=0.95\textwidth]{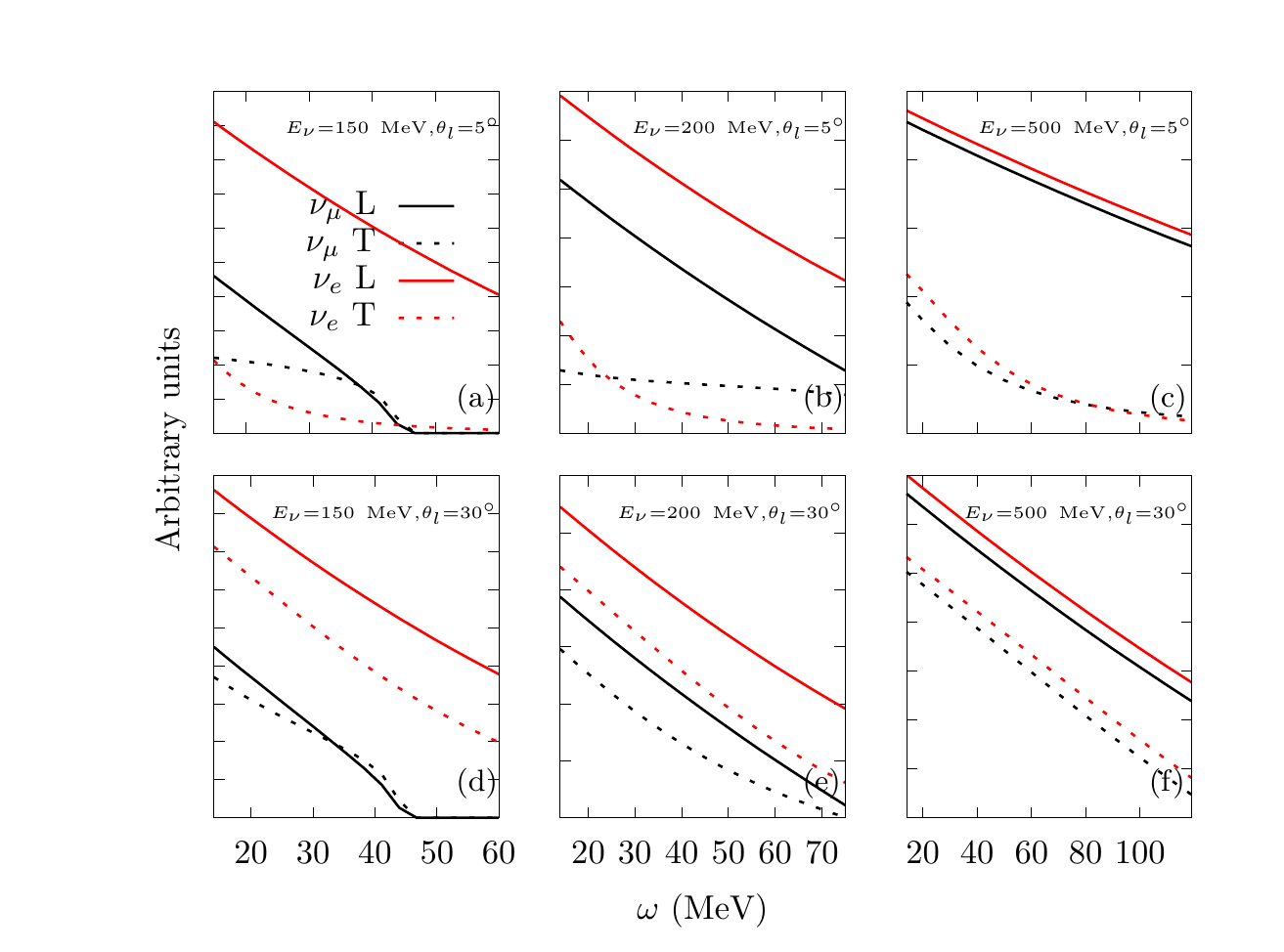}
\caption{The Coulomb-Longitudinal  and transverse (T + T') leptonic factors, containing all the direct dependence on lepton kinematics.}
\label{fig:v_facs}
   \end{center}
\end{figure}

The lepton kinematic factors which are combined with the nuclear responses are depicted in Fig.~\ref{fig:v_facs} which include the Mott-like prefactor (see Eq.~(\ref{eindelijk})), such that they contain all the dependence of the cross section on lepton kinematics.
For most kinematic regions the electron is indeed preferred in the leptonic vertex. 
Around the muon threshold and for very forward scattering angles however, the muon has a non-negligible transverse contribution, while for the lighter electron the transverse contribution is almost zero.
For larger energies the transverse factors become comparable, and small compared to the longitudinal contribution for forward scattering angles.

\begin{figure}
   \begin{center}
\includegraphics[width=0.95\textwidth]{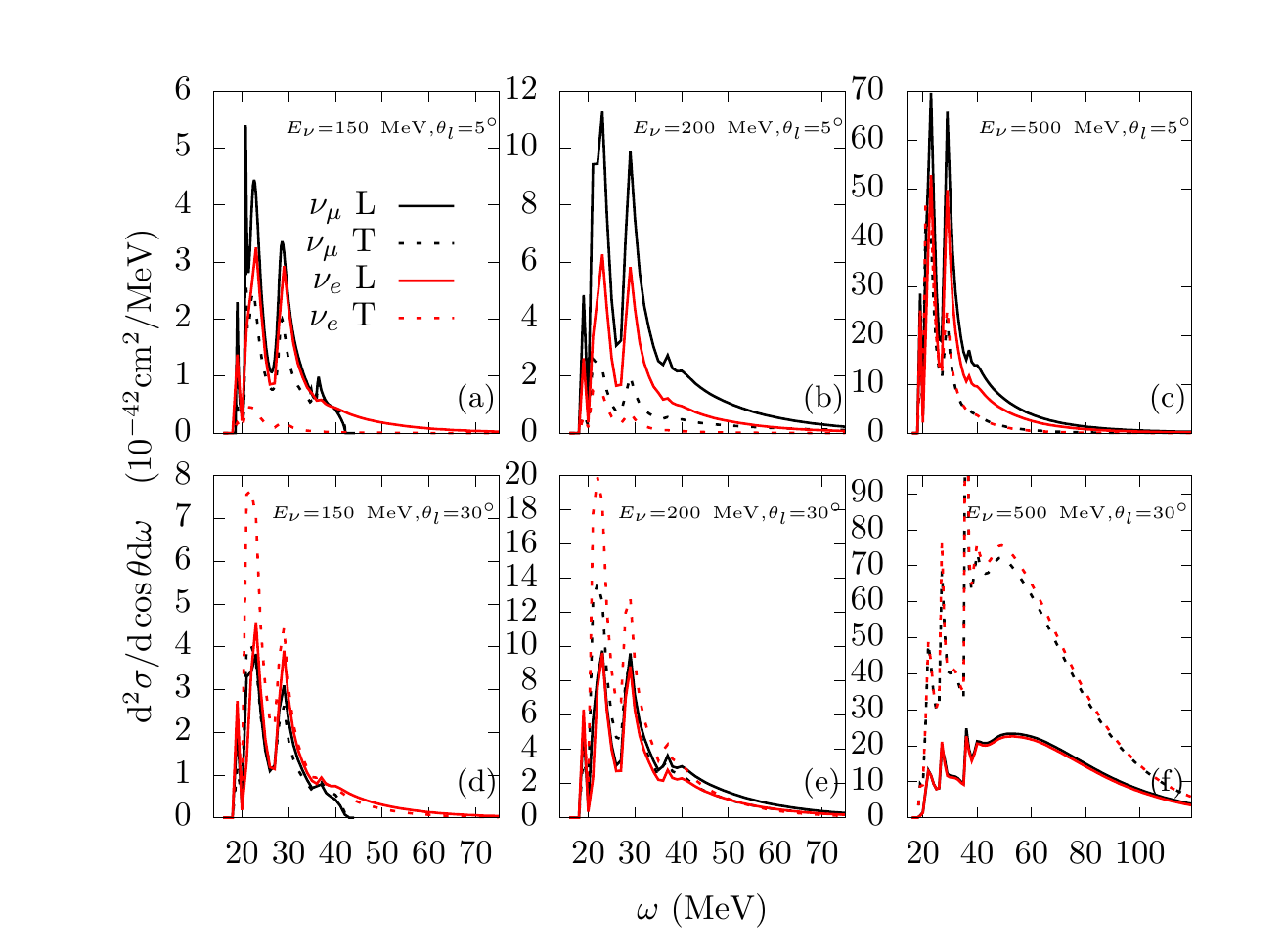}
\caption{Longitudinal (CC+LL+CL) and transverse (T + T') cross sections for the same kinematics as presented in Fig.~\ref{fig:v_facs}.}
  \label{fig:Resp} \end{center}
\end{figure}

From these considerations, one would generally expect larger cross sections for the electron neutrino, but once the nuclear responses are taken into account this picture can actually be reversed.
This is illustrated in Fig.~\ref{fig:Resp}, comparing the longitudinal and transverse contributions to cross sections for muon and electron neutrinos in a CRPA calculation. 
For low incoming energies and scattering angles,  the transverse contribution indeed adds additional strength to the muon-neutrino induced cross section, while it is very small for electron neutrinos. On top of that, the longitudinal contribution is actually larger for the $\nu_{\mu}$ than for incoming $\nu_e$.  This effect finds its origin in the dependence of the nuclear responses on energy and momentum transfer $(q,\omega)$, where for identical values of energy transfer, energy momentum conservation dictates enhanced $q$-values for muons with the larger $m_l$ at forward lepton scattering angles $\theta_l$ :
\begin{equation}
q =\sqrt{ E_{\nu}^2 + P_l^2 -2\cos\theta_l E_{\nu} P_l } \approx E_{\nu} - \sqrt{\left(E_{\nu} - \omega \right)^2 - m_l^2 }.
\end{equation}
At larger energies,  the $\nu_e$ and $\nu_{\mu}$ transverse cross sections become comparable in magnitude, and the longitudinal cross section is the dominating cause of  slightly larger muon-neutrino induced cross sections.

For low energies and larger scattering angles,  the electron neutrino gets a larger transverse contribution, but still the longitudinal muon-induced cross section remains larger than could be expected from the lepton kinematics alone. Only when the neutrino energy and scattering angle become large enough one sees that the cross sections are comparable to the values that can be expected from the lepton factors.

\section{Coherent versus quasi-elastic scattering at low energies}
At very low neutrino energies, inelastic processes are outnumbered by coherent scattering reactions, in which the neutrino scatters off the nucleus as a whole, 
without resolving the individual nucleons. Since this process is cleanly predicted by the Standard Model, it provides an attractive avenue with which we can constrain non--standard interactions.  The lack of detectable reaction products make experimental studies of the process challenging, as these have to rely on measurements of the (small) recoil energies of the target nuclei.  Recently, the COHERENT collaboration managed to measure for the first time the recoil signal produced by coherent scatterings of a nuclear target \cite{Akimov:2017ade}.

The coherent reaction mechanism however has the advantage that the cross section is relatively large, and dominates the  inelastic neutrino-nucleus scattering processes for incoming energies up to a few tens of MeVs. This makes the coherent process important for astrophysical neutrinos, where the large cross sections make it an important instrument for the transfer of energy from the neutrino to the surrounding material. In particular, this is the case for supernova neutrinos, both for their interactions within the collapsing and exploding star core as for their detection on earth.
 For coherent elastic neutrino-nucleus scattering (CE$\nu$NS), the differential cross section is given by the expression
\begin{equation}\label{eq:cevnsdifft}
\frac{\mathrm{d}\sigma}{ \mathrm{d}T} = \frac{G_F^2}{4\pi} M_A \left(1-\frac{T}{E_i}-\frac{M_A T}{2E_i^2}\right){Q_W^2}F(-2M_AT)^2,\label{sevens}
\end{equation}
where $M_A$ is the mass of the struck nucleus, $E_i$ the incoming neutrino energy, $T$ the nuclear recoil, $Q_W$ the weak charge $(1-4\sin{\theta_W}^2)Z - N$.  $F(Q^2)$ is the elastic form factor
\begin{equation}
  F(Q^2)=\frac{4\pi}{Q_W} \int \left( (1-4\sin^2\theta_W)\rho_p(r)-\rho_n(r)
    \right)\frac{\sin(qr)}{qr}r^2dr,
\end{equation}
with $\rho$ the normalized nucleon distributions.  In our results these are obtained using Hartree-Fock nucleon wave functions, RPA corrections are small.
 Other prescriptions for the elastic form factor exist, see e.g. Refs.~\cite{Cadeddu:2017etk,AristizabalSierra:2019zmy}.
  
The highest possible recoil energy at a given incoming neutrino energy $E_i$ is
\begin{equation}\label{eq:tthres}
T_{MAX} =  \frac{2E_i^2}{2E_i + M_A}.
\end{equation}
\begin{figure}
   \begin{center}
\includegraphics[height=.4\textheight,scale=0.5]{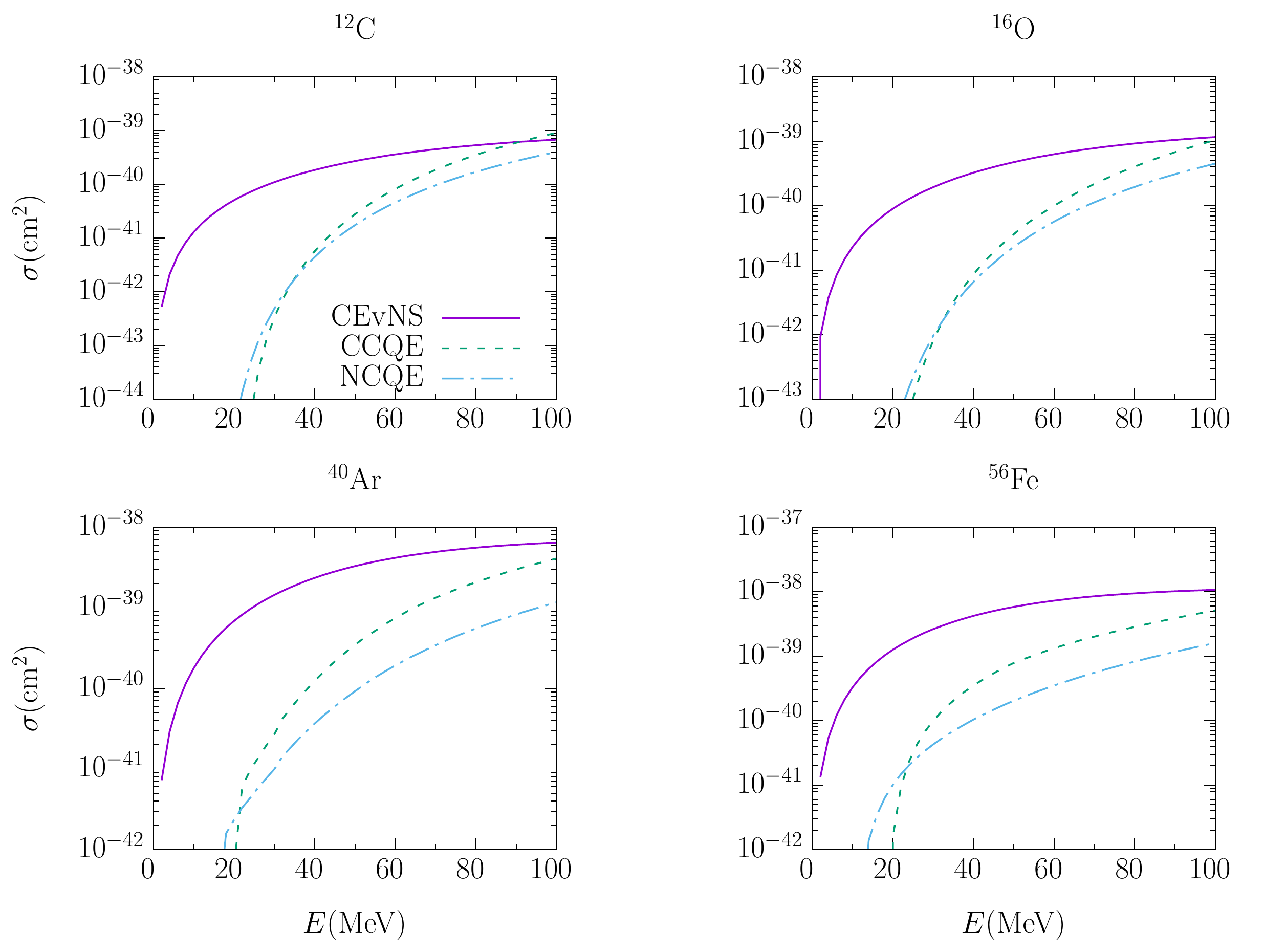}
\caption{CE$\nu$NS total cross section compared to neutrino-induced CCQE and NCQE processes for several nuclei as a function of neutrino energy.}
  \label{coh1} \end{center}
\end{figure}
\begin{figure}
   \begin{center}
\includegraphics[scale=0.5]{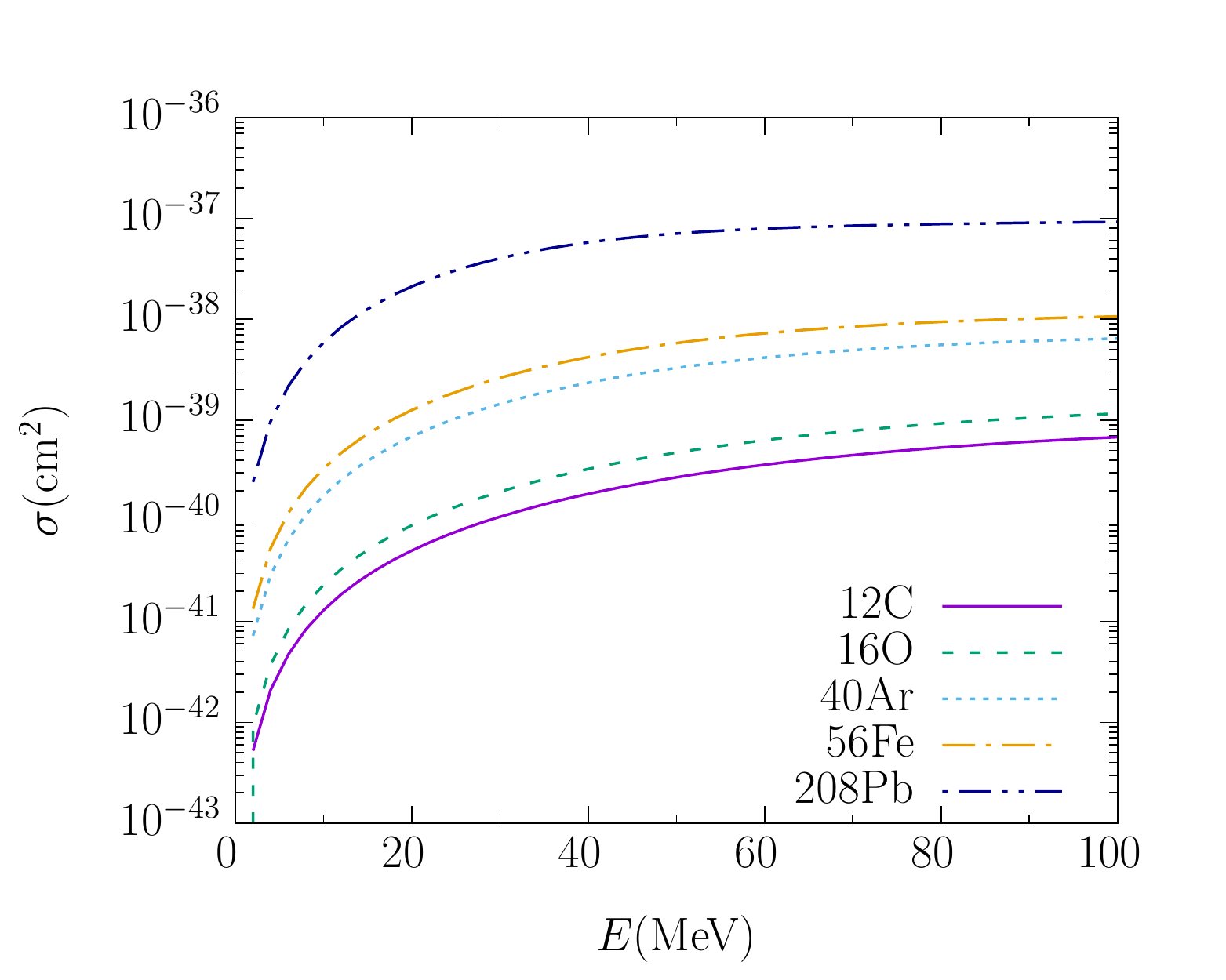}
\caption{CE$\nu$NS total cross sections for several nuclei as a function of neutrino energy.}\label{coh2}
   \end{center}
\end{figure}
\begin{figure}
   \begin{center}
\includegraphics[scale=0.5]{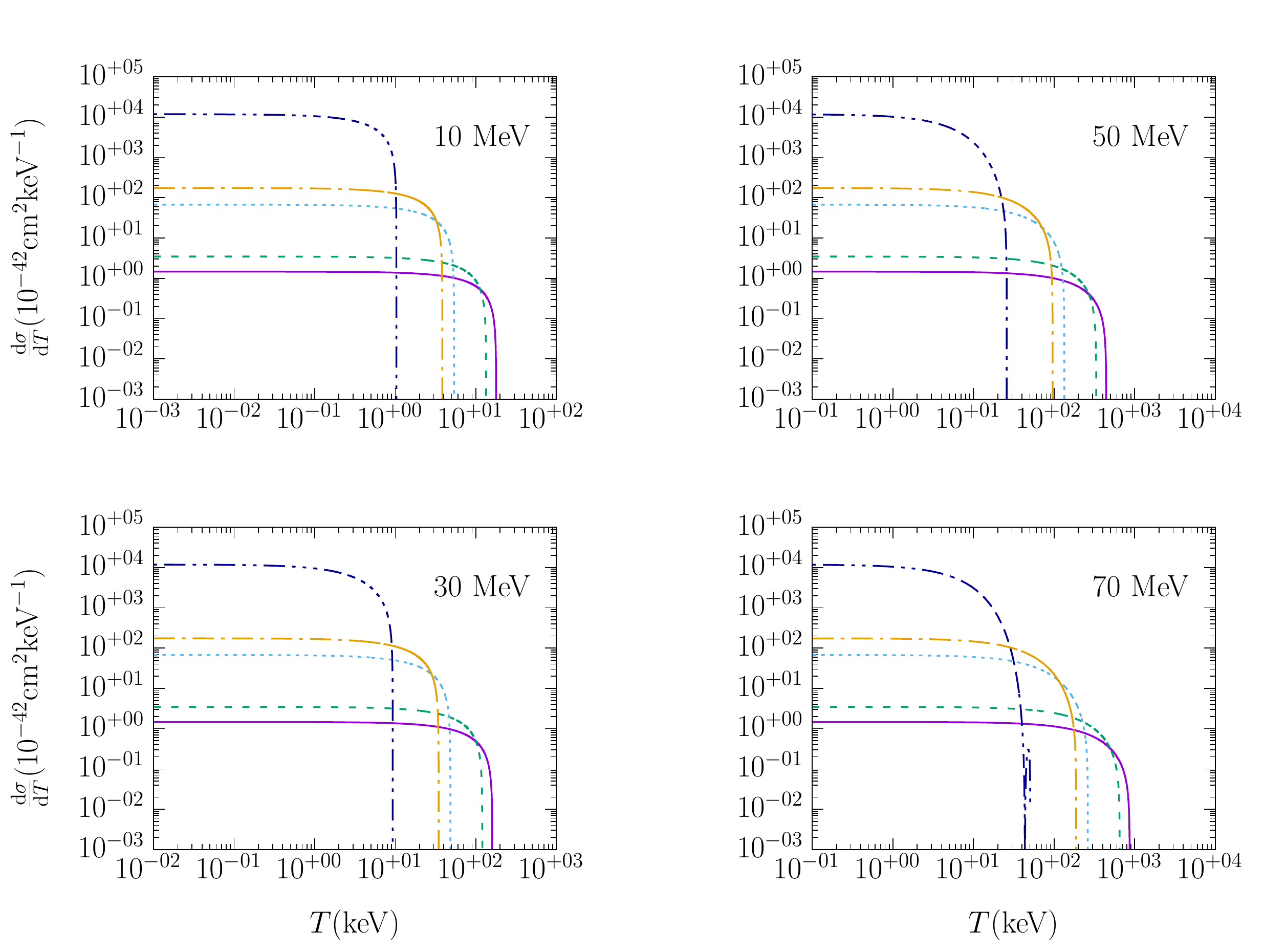}
\caption{CE$\nu$NS differential cross sections as a function of nuclear recoil $T$ for several nuclei for several neutrino energies.  Same key employed as in Fig.~\ref{coh2}}\label{coh3}
   \end{center}
\end{figure}

Since CE$\nu$NS is important for an energy range of roughly $E_\nu \leq$ 100 MeV, and since the rest mass of nuclei is of the order GeV, one can readily appreciate that the kinematically allowed values of $T$ are quite low, in the order of keVs. In particular, the heavier the target nucleus, the smaller its recoil.

To demonstrate the CE$\nu$NS  dominance, we compare it to neutrino-induced quasi-elastic processes with calculations performed in the CRPA framework for several nuclei in Fig.~\ref{coh1}. In general, the total cross section scales strongly as the nucleus increases in mass, roughly proportionate to $N^2$ due to the factor $Q_W^2$, and $\sin{\theta_W}^2 \approx 0.222$. Nevertheless, as the nucleus' size increases, the range of possible nuclear recoils becomes prohibitively small. While at 30 MeV, $^{12}$C is capable of getting a recoil of several 10s of keV, $^{208}$Pb can at most get 9 keV. These effects are  shown in Figs.~\ref{coh2} for total cross sections and \ref{coh3} as a function of recoil for several incoming energies and various targets.

\section{Summary}
We have discussed electroweak interactions between low-energy neutrinos and atomic nuclei. This topic is important for the study of e.g.~supernovae dynamics, with several experiments, in the past, present  and future, dedicated to this line of research, as well as for the low $\omega$ component of reactions with higher incoming neutrino energies.  A thorough theoretical modeling is crucial in this regime. One must account for the outgoing lepton's Coulomb interaction with the residual nucleus, as well as the collectivity present in the nuclear response induced by long--range correlations. Ab-initio, shell-model and RPA-based models offer suitable frameworks in this regard. As we have shown, the peculiarities of low-energy processes  have further implications as illustrated in our discussions on the topics of energy reconstruction and the differences between events induced by electron neutrinos and muon neutrinos. Besides inelastic events, CE$\nu$NS, the elastic scattering of neutrinos off nuclei is also relevant in this energy regime.

\section{Acknowledgments}
This  work  was  supported  by    the  Research Foundation Flanders (FWO-Flanders), and by the
Special Research Fund,  Ghent University.   The computational resources (Stevin\
 Supercomputer Infrastructure)
and services used in this work were provided by Ghent
University,  the  Hercules  Foundation  and  the  Flemish
Government.

\section{Bibliography}

\bibliographystyle{iopart-num.bst}

\bibliography{biblionils}

\end{document}